\documentclass[manuscript]{aastex61}
\usepackage{xspace}
\usepackage{rotating}
\usepackage{gensymb}
\usepackage{multirow}
\usepackage{float}
\usepackage{enumerate}
\usepackage{graphicx}
\usepackage{color}

\newcommand{\enig}{\texttt{\detokenize{enigma_1189}}\xspace}
\newcommand{\enigthree}{\texttt{\detokenize{enigma_1271}}\xspace}
\newcommand{\enigfour}{\texttt{\detokenize{enigma_1266}}\xspace}
\newcommand{\minion}{\texttt{\detokenize{minion_1016}}\xspace}
\newcommand{\astr}{\texttt{\detokenize{astro_1016}}\xspace}

\newcommand{\chtt}{{$C_{H<22}$}\xspace}

\newcommand{\meter}{\,\mathrm{m}}
\newcommand{\km}{\,\mathrm{km}}
\newcommand{\au}{\,\mathrm{au}}

\shorttitle{High-fidelity Simulations of LSST's NEO Search Performance}
\shortauthors{Vere\v{s} and Chesley}

\begin{document}

\title{High-fidelity Simulations of the Near-Earth Object Search Performance of the Large Synoptic Survey Telescope}

\correspondingauthor{Peter Vere\v{s}}
\email{peter.veres@cfa.harvard.edu}

\author{Peter Vere\v{s}}
\altaffiliation{Present address: {\em Minor Planet Center, Harvard-Smithsonian Center for Astrophysics, 60 Garden Street, Cambridge, MA 02138}}
\affil{Jet Propulsion Laboratory, California Institute of Technology, 4800 Oak Grove Drive, Pasadena, CA, 91109}

\author{Steven R. Chesley}
\affiliation{Jet Propulsion Laboratory, California Institute of Technology, 4800 Oak Grove Drive, Pasadena, CA, 91109}

\begin{abstract}

We perform high fidelity simulations of a wide-field telescopic survey searching for Near-Earth Objects (NEO) larger than 140\,m, focusing on the observational and detection model, detection efficiency and accuracy. As a test survey we select the Large Synoptic Survey Telescope. We use its proposed pointings for a 10-year mission and model the detection of near-Earth objects in the fields. We discuss individual model parameters for magnitude losses, vignetting, fading, asteroid rotation and colors, fill factor, limiting magnitude, rate of motion, field shape and rotation and survey patterns and we assess results in terms of the cumulative completeness of the detected population as a function of size and time. Additionally, we examine the sources of modeling uncertainty and derive the overall NEO population completeness for the baseline LSST survey to be $55\pm5$\% for NEOs with absolute magnitude brighter than 22. Including already discovered objects and ongoing surveys, the population completeness at the end of the LSST baseline survey should reach $\sim 77\%$.

\end{abstract}


\section{Introduction} \label{sec:intro}

The number of currently discovered asteroids is about 730,000, the vast majority of which are found in the main asteroid belt. The size of the asteroid catalog has sky-rocketed with the dawn of automated, CCD-based asteroid surveys  only two decades ago. The first asteroid discovered, Ceres, was discovered in 1801 and since 2006 has been classified as a dwarf planet. In 1900, only 545 asteroids where known. The first Near-Earth Object (NEO), 433 Eros, was discovered two years earlier. In 1992 recognition of NEOs as a potential threat to Earth led the US Congress to mandate that NASA discover 90\% of NEOs larger than $1\km$ within 10 years. Thus the Spaceguard survey, aimed at the discovery and study of NEOs, was established. The first generation of dedicated asteroid surveys was developed under this initiative, e.g., Spacewatch \citep{2006DPS....38.5807M}, LINEAR \citep{2000Icar..148...21S}, NEAT \citep{1999AJ....117.1616P}, LONEOS \citep{1999BAAS...31.1091K}. As a direct result of these efforts, the number of NEOs increased from 649 in 1998 to over 5000 in 2008, and as a byproduct, other small body populations of the Solar System were discovered as well, ranging from main belt asteroids, to Jupiter Trojans, Centaurs, Kuiper belt objects and comets. With new technologies becoming available, and new challenges such as impact hazard mitigation of smaller NEOs, potential commercial activities like asteroid mining, asteroid sample return, and asteroid redirection, as well as crewed mission concepts to an NEO, a new target was set to discover 90\% of NEOs larger than 140\,m.  Now the next generation of Solar System surveys has emerged, notably, Pan-STARRS \citep{2002SPIE.4836..154K,2004SPIE.5489..667H} and the Catalina Sky Survey \citep{2003DPS....35.3604L,2012DPS....4421013C}, which are now dominating minor planet discoveries, which reached 15,000 in 2016. Meanwhile, NEOWISE \citep{2010AJ....140.1868W} is a space-based infrared telescope that is finding many asteroids, and providing diameter estimates for even more. However, the current NEO completeness for 140\,m and larger objects is only about 30\% and the currently operating search programs are not able to achieve the 90\% goal in the next few decades. Therefore, a number future projects have been proposed and are in varying stages of development, such as the Zwicky Transient Factory \citep{2014SPIE.9147E..79S}, NEOCAM \citep{2015DPS....4730801M} and the Large Synoptic Survey Telescope \citep{jones2017}.

Survey discovery rates and statistics, along with information on the sky coverage and telescope sensitivity, are widely used for determination of the population counts and statistical distribution of orbital elements. For instance, \citet{2002Icar..156..399B, 2004Icar..170..295S,2011ApJ...743..156M, 2015Icar..257..302H,2016Natur.530..303G,2017Icar..284..114S} utilized discovery statistics and the telescopic biases to derive NEO population properties. This reverse engineering and debiasing led to estimates of the size-frequency distribution of a given population and is a key source of information about the current state of the Solar System inventory and an input for modeling the origin and evolution of planets and minor bodies. These models have uncertainties, mostly originating in errors in determination of the absolute magnitude $H$ of asteroids and poorly understood search and detection efficiencies of individual surveys. 

The opposite approach allows prediction of discovery rates for current and future telescopes. This requires knowledge of the key parameters of the telescopes and their detection pipelines. Even though simulations on discovery rates have been performed for Solar System surveys, the limiting parameters, errors or uncertainties are often omitted and their effects are not discussed or fully disclosed in literature. Without a complete description of the input parameters and their effect on the detection efficiency, a useful comparison of survey simulations done by different authors is challenging and often infeasible.

Traditionally, ``discovery" means not just the first observation of an asteroid but also its confirming astrometry from the following nights, until a provisional orbit is obtained. Future large-scale, deep, all-sky surveys, like LSST, will not have the luxury of follow-up confirmation and will depend on self-follow-up capabilities. This will remove the difficult-to-model human factor from the process of discovery. Moreover, automated pipelines and processing are now more frequently responsible for object detection and reporting, which allows a more exact description of the input parameters and detection thresholds. Surveys also detect objects already discovered and have to identify them in the pipeline.

Our work focused on the NEO detection and discovery rates and its variations as a function of the selected observational and detection parameters.  We studied a broad range of effects, including the magnitude variations due to vignetting, asteroid colors and light curves, the rate of motion of asteroids, limiting magnitudes, the probability of an object being detected when near the faint limit, the shape and rotation of the field, fill factor, the survey search pattern, and the overall sky coverage. As a case study we consider the Large Synoptic Survey Telescope (LSST), which will start its commissioning in 2020 and its main 10-year mission in 2022. Even though LSST will be mainly focused on understanding and studying Dark Matter and Dark Energy, its all-sky coverage, large aperture and the largest CCD camera in the world make it an effective Solar System cataloging tool that is predicted to increase the number of known Main belt asteroids to 5 million, NEOs to 100,000 and Trans-Neptunian Objects to 40,000 \citep{2016IAUS..318..282J}. We used the Moving Object Processing System \citep[MOPS,][]{2013PASP..125..357D} to simulate the detections, and for linking of detections into tracklets. MOPS and its variations have been used to process real data of Pan-STARRS, DECAM \citep{2015AJ....150..150F}, and NEOWISE, and will be used by LSST. MOPS computes ephemerides of input synthetic orbits with defined $H$ for a set of telescope pointings, provides visible detection lists, and submits it to the next stage to create {\em tracklets}, which are the single-night detections of a candidate moving object. MOPS has a capability of linking inter-night tracks and derived orbits, however, it has never been tested in such a high-density detection and tracklet environment like LSST. Thus, our work assumes that the linking to orbits will work on the available collection of tracks. In a separate paper \citep{2017Veres_2}, we consider the linking efficiency problem for LSST with MOPS.

The primary survey performance metric of this work is the cumulative completeness of the detected NEO population in detections, tracklets and available 3-night tracks as a function of $H$, particularly for $H<22$ NEO (\chtt), and as a function of time. We also study the effects of using two input NEO populations, detection efficiency of Potentially Hazardous Asteroids (PHA) and compare our results with previous work.

\section{Large Synoptic Survey Telescope}\label{sec:LSST}

LSST \citep{Ivezic2014} is being built atop Cerro Pach\'{o}n in Chile and is funded by the National Science Foundation, the Department of Energy and private contributors. It is a next generation optical and near-infrared all-sky survey telescope with the mission of sparse sampling of the sky in short exposure times (Table~\ref{tab.LSSTtab}). A single 30-second exposure with a 9.6 square degree field of view will achieve 24.5 magnitude depth in r-band. LSST will scan 6,000 square degrees per night. Because of the unprecedented quantities of data and LSST's real-time alerting requirements, the detection, processing and data handling will be fully automated.

\begin{figure}[htb]
  \centering
  \epsscale{0.7}
\plotone{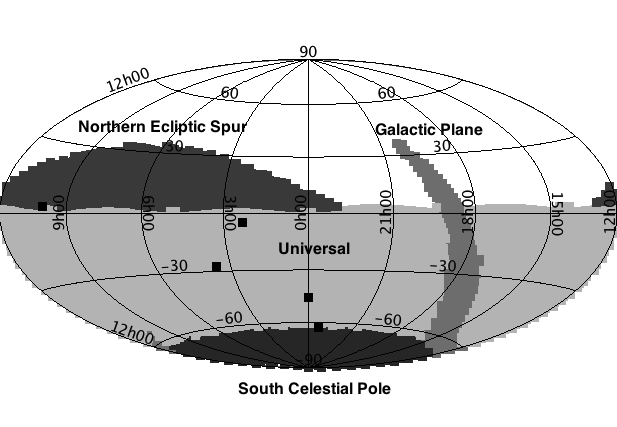}
     \caption{10-year sky coverage of the \enig survey. Note that the Deep Drilling proposal covers only a few (black) fields. Most of the time is spend in the Universal proposal that covers most of the available sky.}
    \label{fig.coverage}
\end{figure}

\begin{figure}[htb]
  \centering
    \epsscale{0.5}
    \plotone{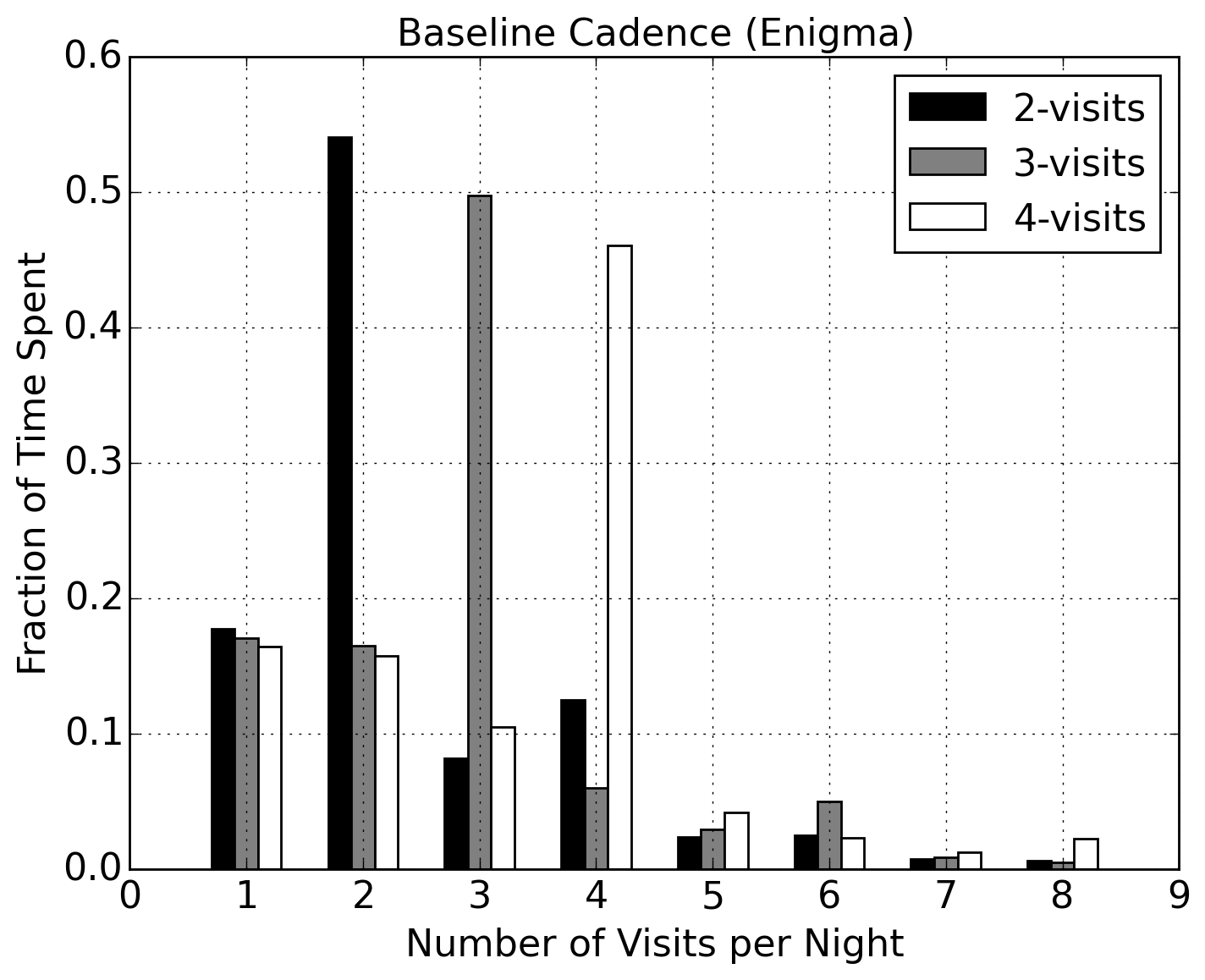}
     \caption{Baseline surveys optimized for 2, 3 and 4 visits and the histogram of time spent on fields visited N-times per night.}
    \label{fig.234visits}
\end{figure}

Scheduling of the visited fields is done through the Operations Simulator \citep[OpSim,][]{2014SPIE.9150E..15D}. Fields are visited in five ``proposals", each focusing on different science. Most of the time (85\%) is spent in the Universal proposal that covers the entire Southern Hemisphere, the rest is divided between the coverage of the Galactic Plane, Northern Ecliptic Spur, South Celestial Pole and a few Deep Drilling (DD) fields (Figure~\ref{fig.coverage}). The final survey pattern is not yet selected, but several scenarios have been tested. We used some of the more relevant OpSim runs for our simulations: the old LSST baseline \enig, the new baseline \minion and NEO-enhanced \astr (which we abbreviate from the full simulation designation \texttt{\detokenize{astro_lsst_01_1016}}). Most of the fields are observed twice per night, except for two \enig variations, where three (\enigthree) and four visits  (\enigfour) were targeted (Figure~\ref{fig.234visits}). Some of the fields are only observed once per night (singletons). These were removed from our simulation because they cannot aid discovery. Also, the DD proposal covers only a few individual fixed fields that are visited in an extreme revisit cadence (tens of times per night, one by one) and are not suitable for NEO discovery, therefore, DD was removed as well. The median revisit time between exposures of the same field in a night is 21 minutes and in 94\% of cases the first and last visit are within 2 hours in a night (Figure~\ref{fig.234visits}). Each field is observed in one of the six filters (Table~\ref{tab.surveys_mag}) and at a specific limiting magnitude, seeing, simulated atmospheric conditions and lunar phase. 

\begin{table}[H]
\small
\setlength\tabcolsep{1.5pt}
\begin{center}
\caption{LSST in numbers}
\begin{tabular}{l|c}
\tableline\tableline
Location & Cerro Pach\'{o}n, Chile\\
Primary mirror& 8.4\,m\\
Effective Aperture & 6.67\,m\\
Focal Length&10.3\,m\\
Wavelength & 320-1060\,nm\\
Filters & u, g, r, i, y, z \\
Mount&Alt-Az\\
Field diameter&3.5\,deg\\
Camera & 3.2\,Gigapixel\\
Pixel size & 0.2\,arcsec\\
Exposure time&30\,s\\
Start year & 2022\\
\tableline
\end{tabular}
\label{tab.LSSTtab}
\end{center}
\end{table}

\begin{table}[htb]
\begin{center}
\caption{SNR=5 limiting magnitudes ($m_5$) of \enig OpSim survey and time spent in individual filters.}
\begin{tabular}{c|ccc}
\tableline\tableline
Filter & Average $m_5$ & Max $m_5$ & Time spent (\%)\\
\hline
r&24.09$\pm$0.33&24.91&22.00\\
i&23.34$\pm$0.37&24.33&22.00\\
z&22.28$\pm$0.41&23.71&20.00\\
y&21.56$\pm$0.23&22.21&18.00\\
g&24.38$\pm$0.35&25.24&10.00\\
u&23.09$\pm$0.33&23.99&7.00\\
\tableline
\end{tabular}
\label{tab.surveys_mag}
\end{center}
\end{table}

We generated synthetic detections for NEO and MBA population models (see Section~\ref{sec:populations}) by propagation of the orbits to the epochs of the OpSim fields. The propagation used JPL's small body codes with the DE405 planetary ephemerides, where all planets, plus Pluto and the Moon were perturbing bodies. We did not use any asteroids as perturbers. Only detections inside of the fields of view were saved into preliminary detection lists. The limiting magnitude $m_5$ of the field is defined at a signal-to-noise ratio of SNR=5. Photometric or astrometric errors associated with the astrometry were neglected because the are not relevant to the study approach, which does not include orbit fitting. The following subsections describe the details of the NEO detection modeling approach.

\subsection{Population models}\label{sec:populations}

This work tested two NEO population models. Each model is represented by a set of Keplerian orbits with absolute magnitudes $H$ defined in Johnson's V-band following the size-frequency distribution. 

\begin{figure}[H] 
  \centering
  \epsscale{0.6}
    \plotone{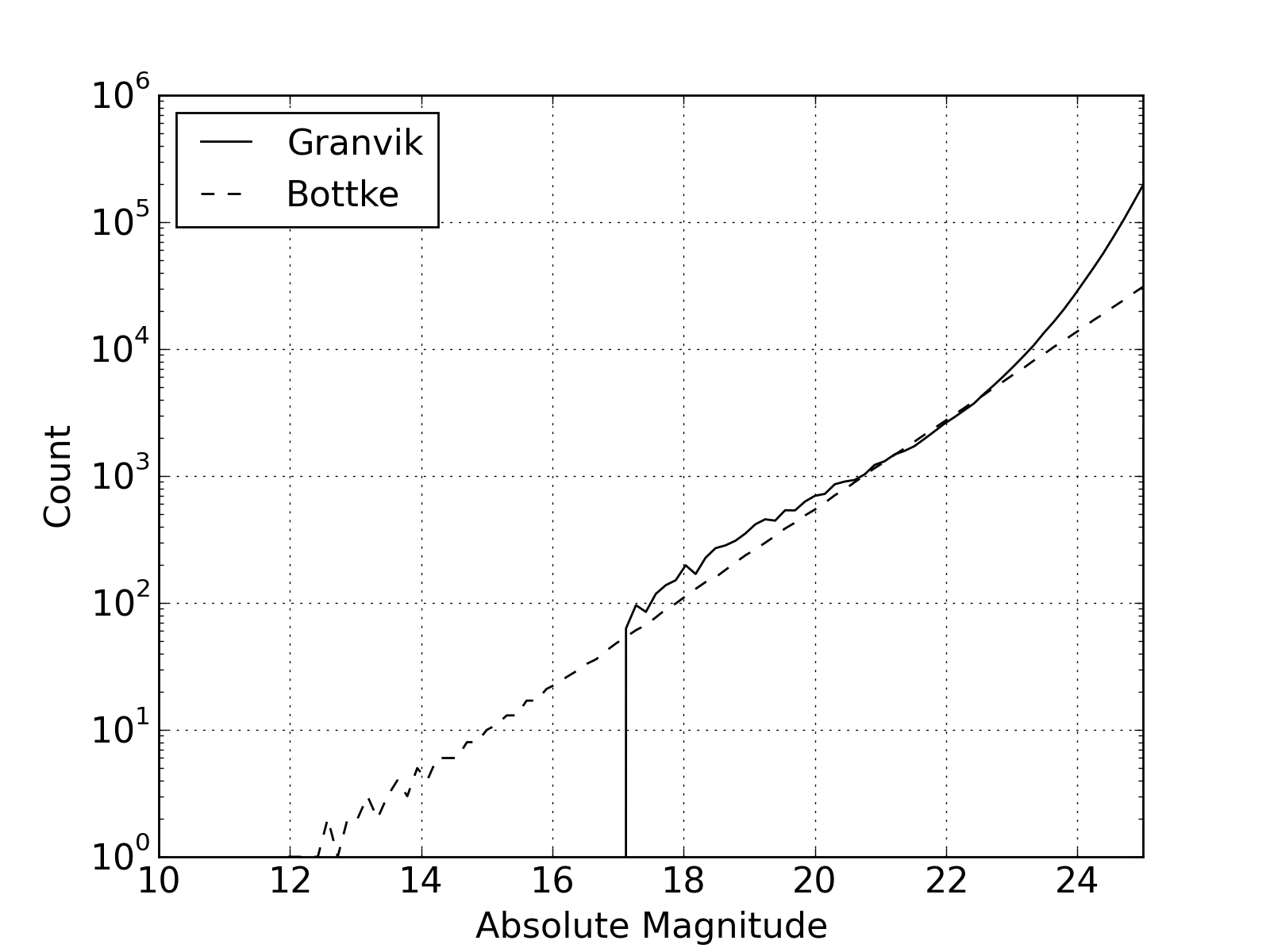}
   \caption{Cumulative distribution of Bottke's and Granvik's NEO populations.}
    \label{fig.NEO_SFD}
\end{figure}

\begin{figure}[H]
  \centering
    \includegraphics[width=0.43\textwidth]{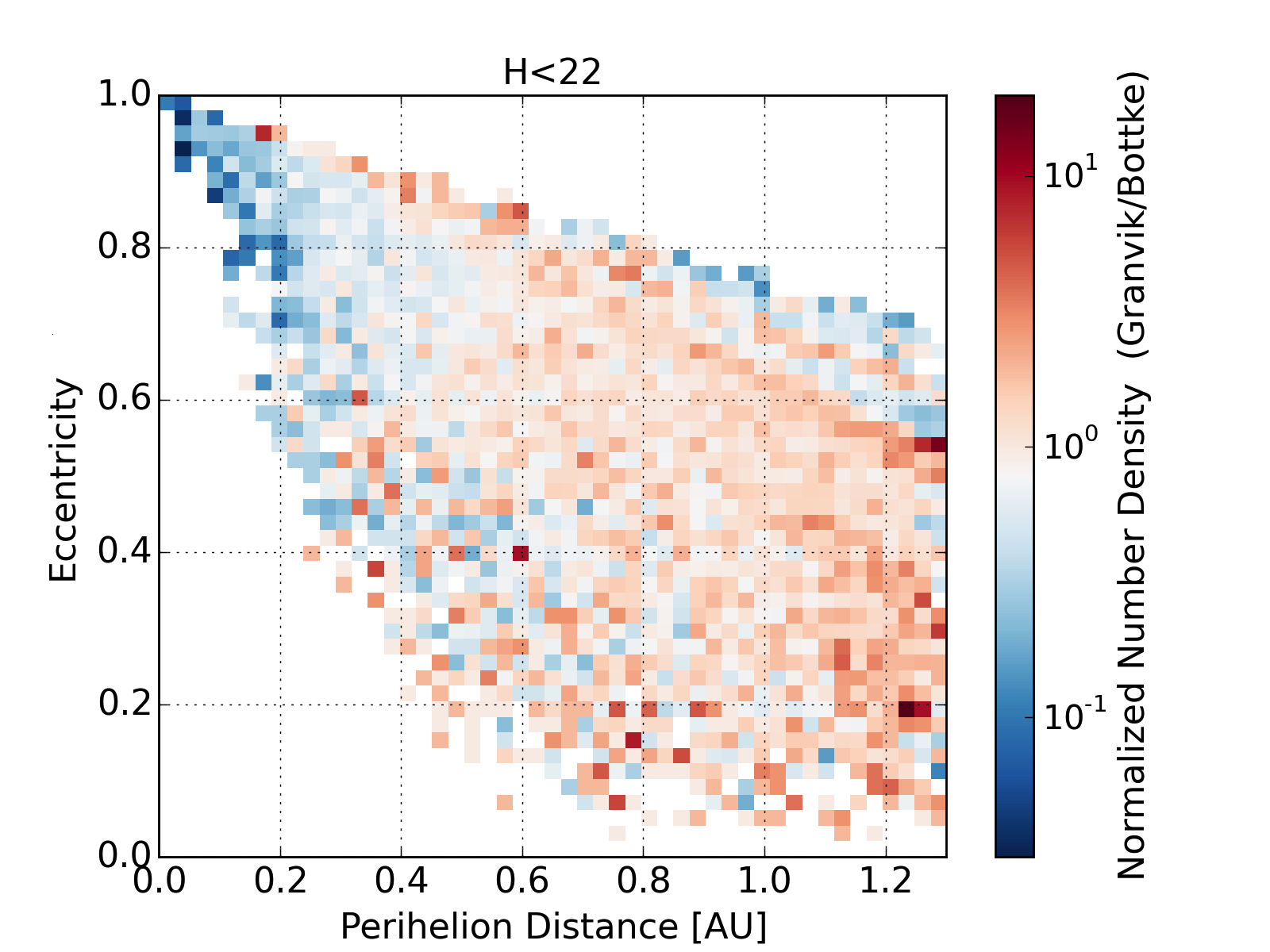}
    \includegraphics[width=0.43\textwidth]{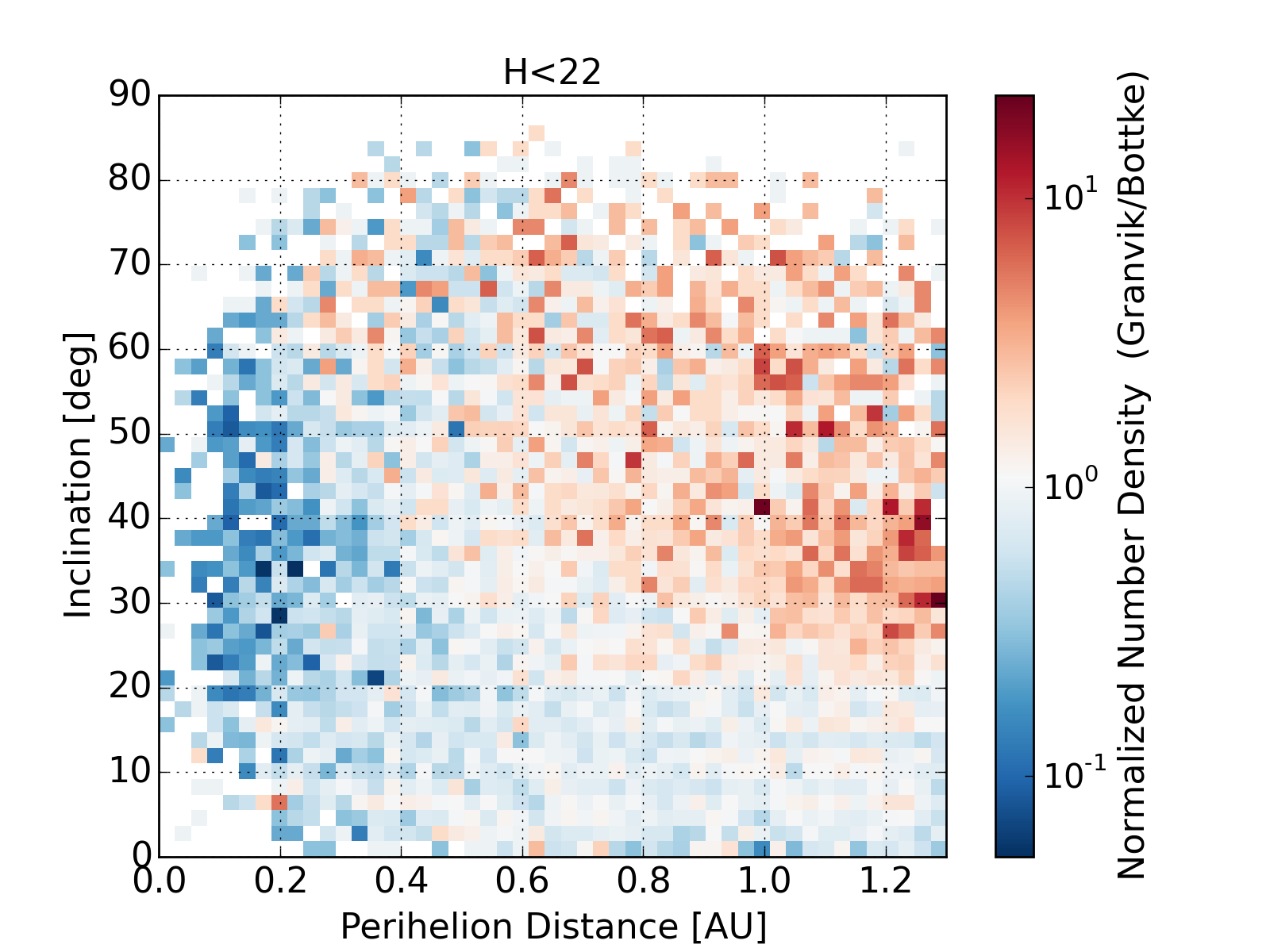}
    \includegraphics[width=0.43\textwidth]{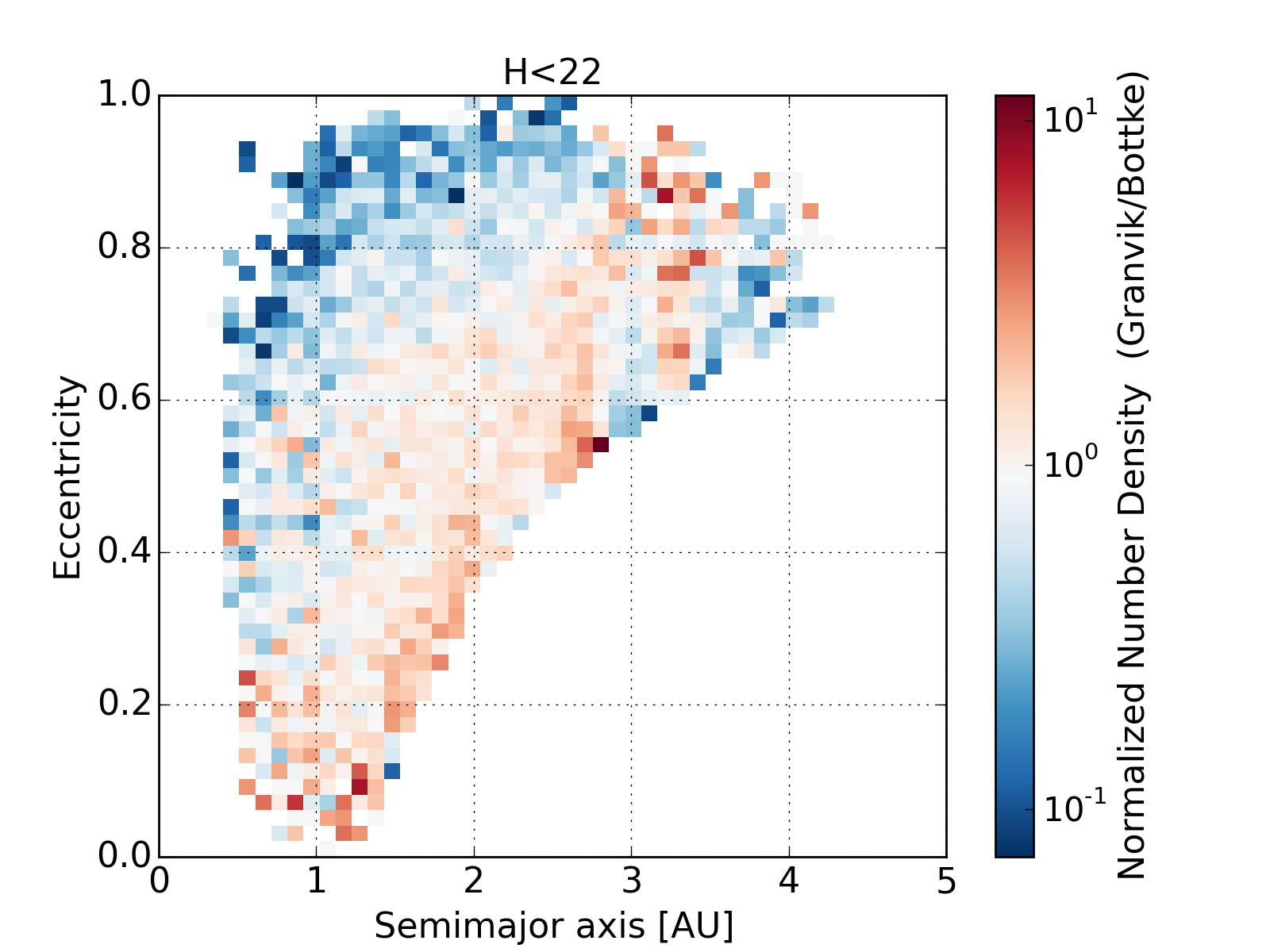}
    \includegraphics[width=0.43\textwidth]{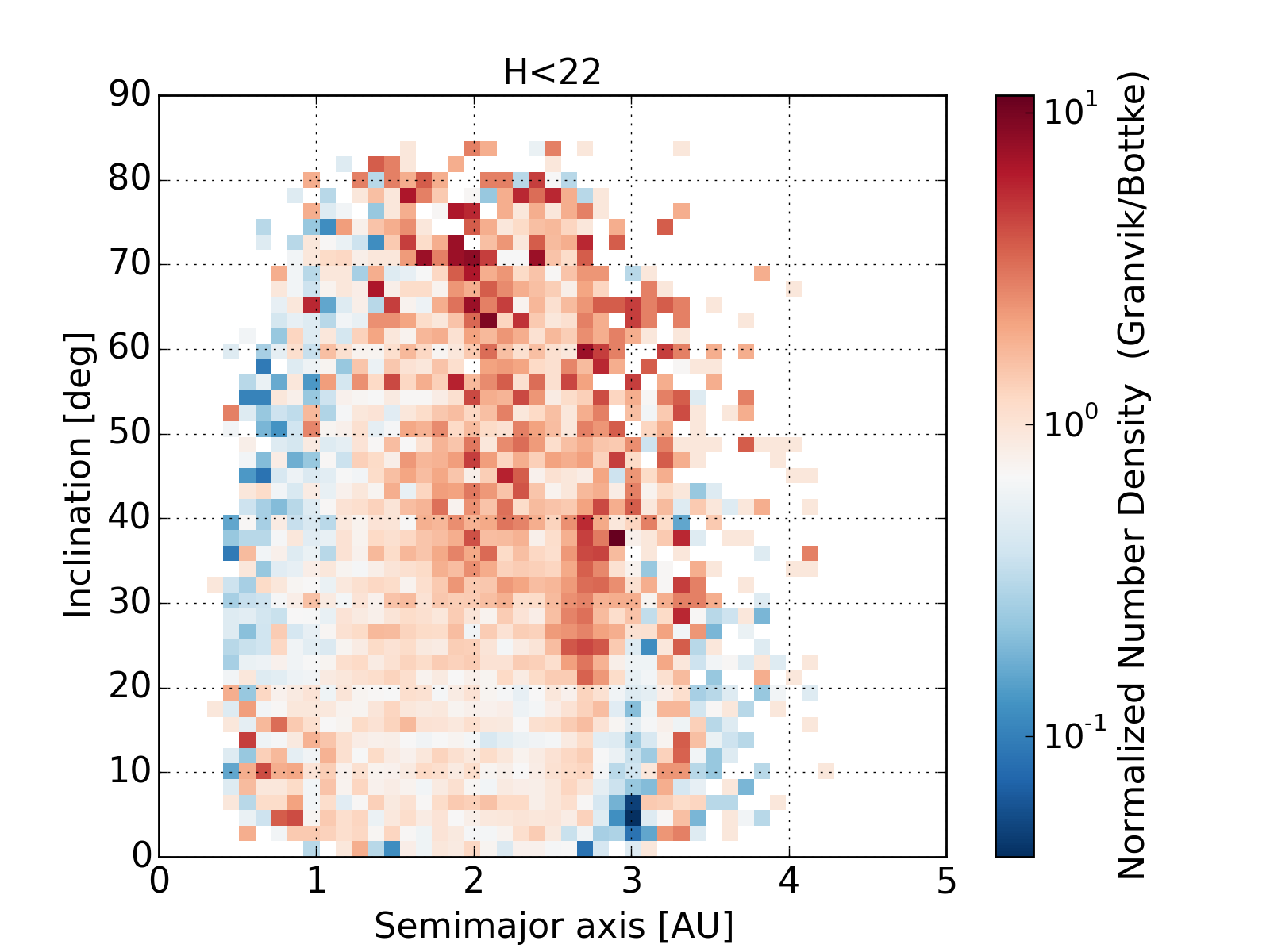}
    \caption{Comparison of orbital element distribution between Bottke's and Granvik's NEO populations. These density plots reflect the ratio of the normalized density between the respective models.}
    \label{fig.NEOdiff}
\end{figure}

The first NEO model that we used is from an earlier work \citep{2002Icar..156..399B, 2011PASP..123..423G}. ``Bottke's'' model consists of 268,896 orbits with $H<25$. A newer model was published by \citet{2016Natur.530..303G} after our work was already underway, which we refer to  as ``Granvik's'' model. It has 801,959 orbits, again down to $H<25$. Even though Granvik's model has three times as many objects, the number of $H<23$ orbits is about the same as in the Bottke's data set (Figure~\ref{fig.NEO_SFD}). On the other hand, Granvik's model lacks objects with $H<17$. For simulations, these objects can be omitted because such bright NEOs are few and mostly discovered. The main difference is in the slope of the distribution for $H>23$, where Granvik's population describes the observational data better than Bottke's early estimate. Bottke's orbital element distribution does not depend on $H$, while Granvik's population does. The slight differences between the distributions of orbital elements are evident in Figure~\ref{fig.NEOdiff}. Compared to Granvik, Bottke predicted more objects at low perihelion distances and larger eccentricities, while Granvik, on the other hand, shows an excess at larger perihelion distance and inclination.

The Potential Hazardous Asteroids (PHAs) are a special subset of the NEOs. They are defined as objects with the minimum orbit intersection distance (MOID) less than $0.05\au$ and $H<22$. $H=22$ corresponds to the diameter of 140\,m when albedo is equal to 0.14. We created a subset of PHAs from both Granvik's and Bottke's populations.

\newpage
\subsection{Focal Plane Model}\label{sec:focalplane}

\begin{figure}[htb]
  \centering
  \plottwo{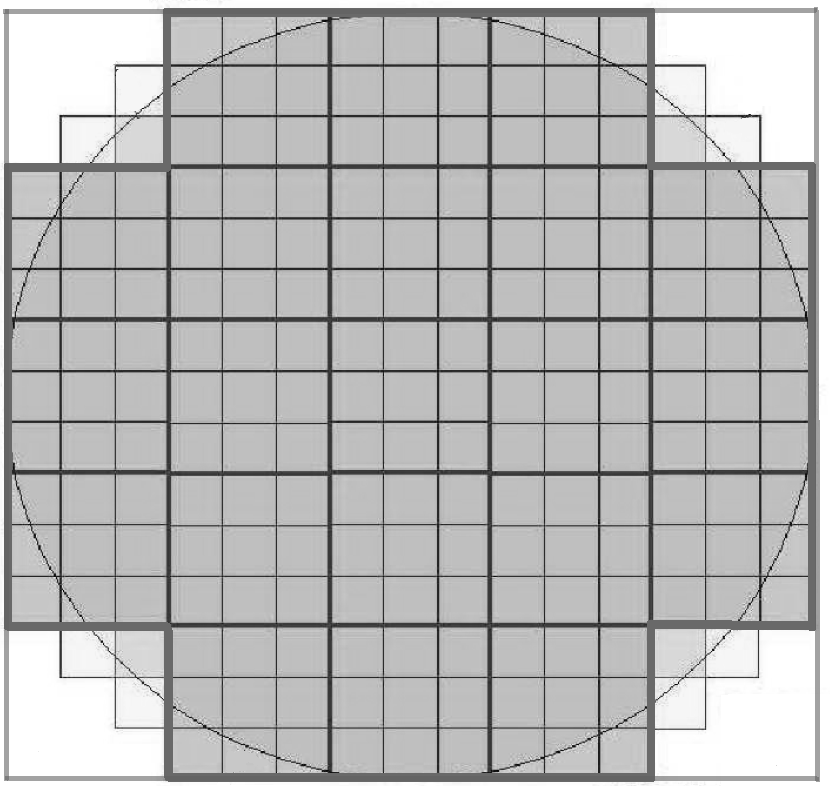}{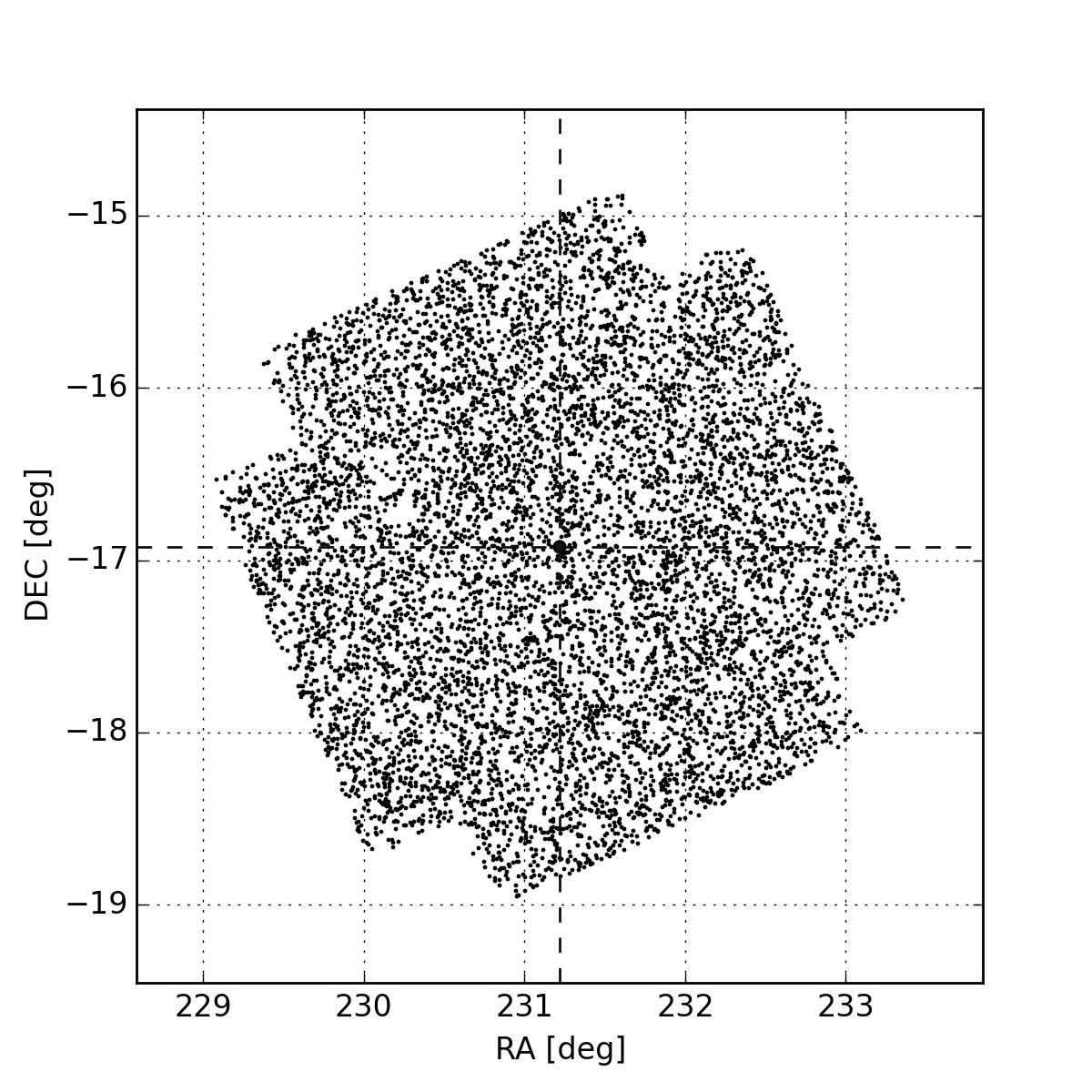}
     \caption{Model of the LSST focal plane. Thick lines represent raft gaps, thin lines chip gaps. Active chips are depicted by gray shading (left). Detections modeled in LSST field rotated by a position angle of 66 deg (right).}
    \label{Fig.focal_plane}
\end{figure}

The focal plane of a large-format survey telescope usually differs from a simple square or a circular shape. LSST's focal plane has a square shape consisting of a 5$\times$5 grid of 25 rafts with the 4 corner rafts removed (Figure~\ref{Fig.focal_plane}). Each raft consists of an array of 9 CCD chips, yielding a total of 189 CCD chips. The detectors are 4096$\times$4096 pixel CCDs, and so the total number of active pixels is 3,170,893,824. The design of the mosaic camera includes chip and raft gaps that do not contain pixels, and therefore, a detection hitting the gap is lost. The shape of the field can be modeled easily and in the simulation we numerically compute the location of each detection with respect to the borders of the camera. Fill factor (the gaps) can be simulated either by exact mapping of the grid to the sky-plane coordinates or by a statistical approach that randomly removes a fraction of detections. The current LSST camera design expects a 90.8\% fill factor.

LSST utilizes an altitude-azimuthal mount and the camera is able to rotate, and thus the fields are not generally aligned with the local RA-DEC frame. In fact, due to desired dithering, each exposure is observed in a randomized field orientation. The change of position angle between two exposures in a night ($\delta_{rot}$) is small. The average $\delta_{rot}$ is $5\degree$ and the median $\delta_{rot}$ is $2.4\degree$.

\subsection{Detection Fading}\label{ssec:fading}

\begin{figure}[H]
  \centering
    \epsscale{1.0}
    \plottwo{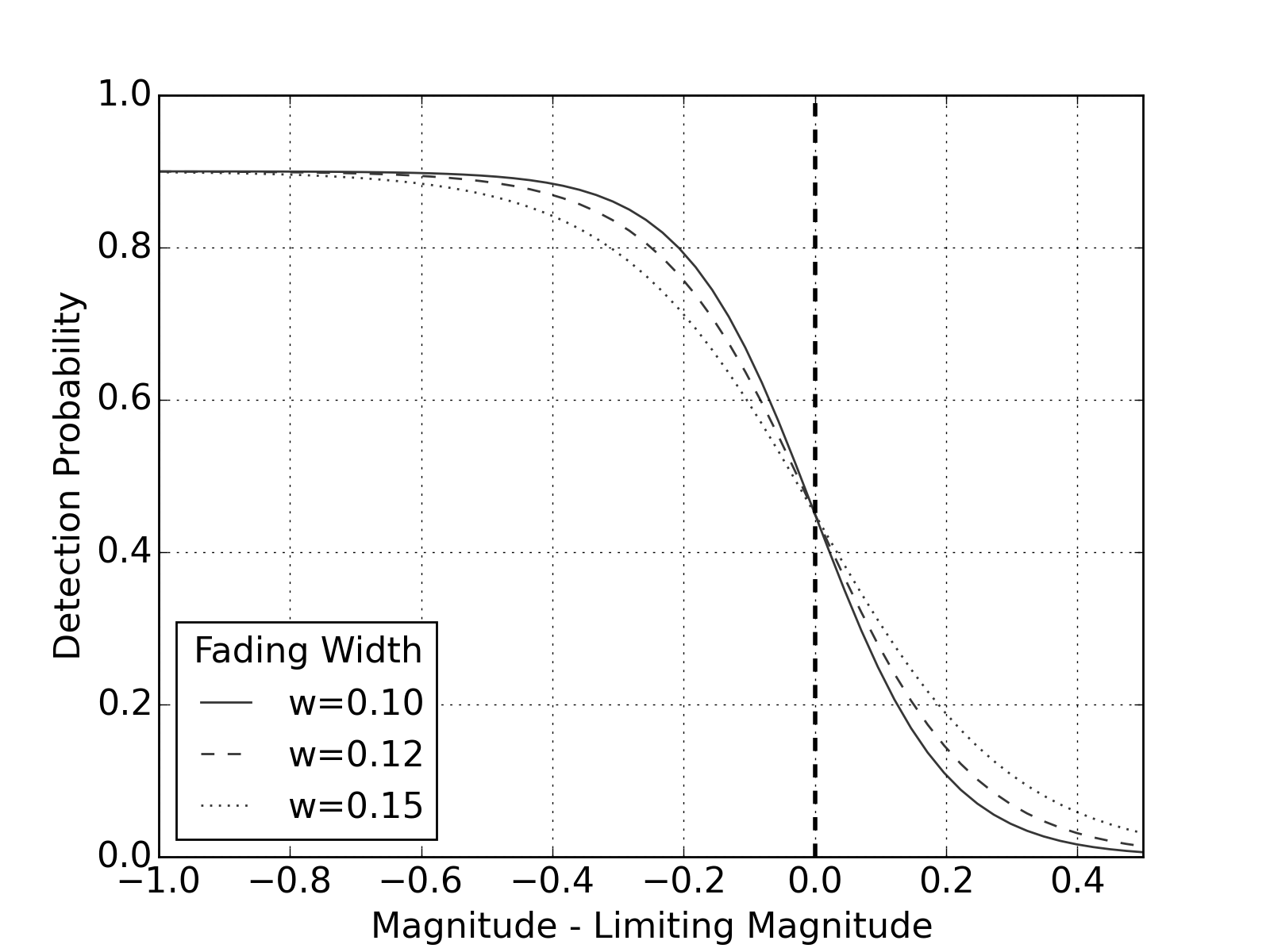}{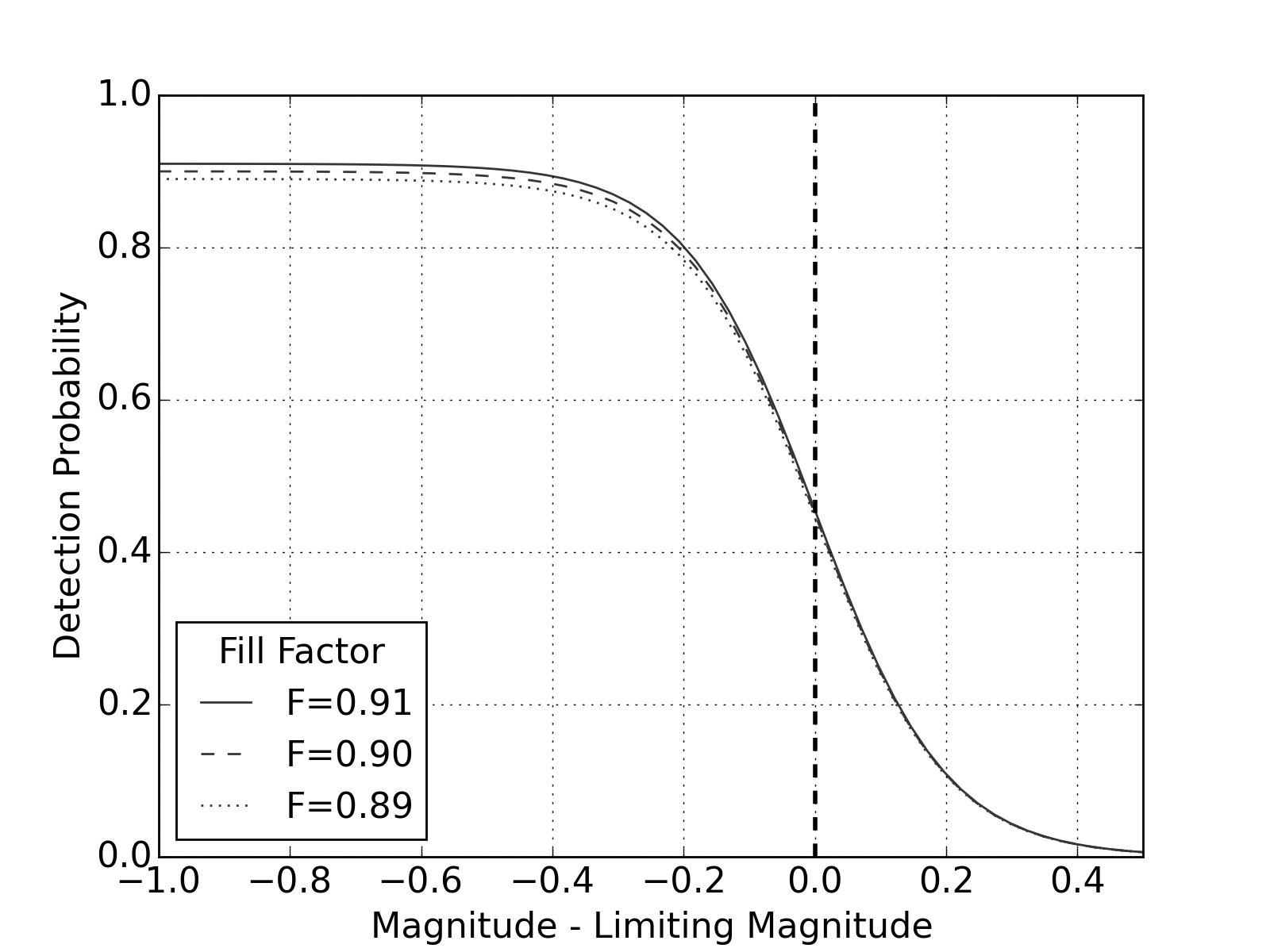}
     \caption{Probability of a target to be detected, with a fixed fill factor of 90\% and variable width (left) and fixed width of $w=0.1$ and variable fill factor (right). The $m_5$ limiting magnitude is equal to a 50\% probability detection, neglecting fill factor.}
    \label{fig.fading}
\end{figure}

In survey simulations the limiting magnitude is often used as a step function cutoff that determines strictly which detections are visible in the simulated field. However, in real surveys, the detection limit is actually better represented by a function that gradually fades near the limiting magnitude to capture the probability of a given detection. The function can be described as
\begin{equation} 
  \label{eq.fading}
	\epsilon(m) 
	   = \frac{F}{1+e^{\frac{m-m_5}{w}}}
\end{equation}
where $\epsilon(m)$ is the probability of the detection, $F$ the fill factor, $m$ the magnitude of the detection, $m_5$ the limiting magnitude of the field and $w=0.1$ the width of the fading function. The fading function is depicted in Figure~\ref{fig.fading}.  Because the limiting magnitude is unique per field, the fading function applies on a per-field basis. The importance of using the fading function lies in the fact that it actually allows detection of asteroids fainter than the $m_5$ limit. Given the brightness distribution of moving objects, this tends to increase the number of detections in the field with respect to a scenario where only the $m_5$ step function was used to determine detection.

\begin{figure}[H]
  \centering
      \epsscale{0.6}
    \plotone{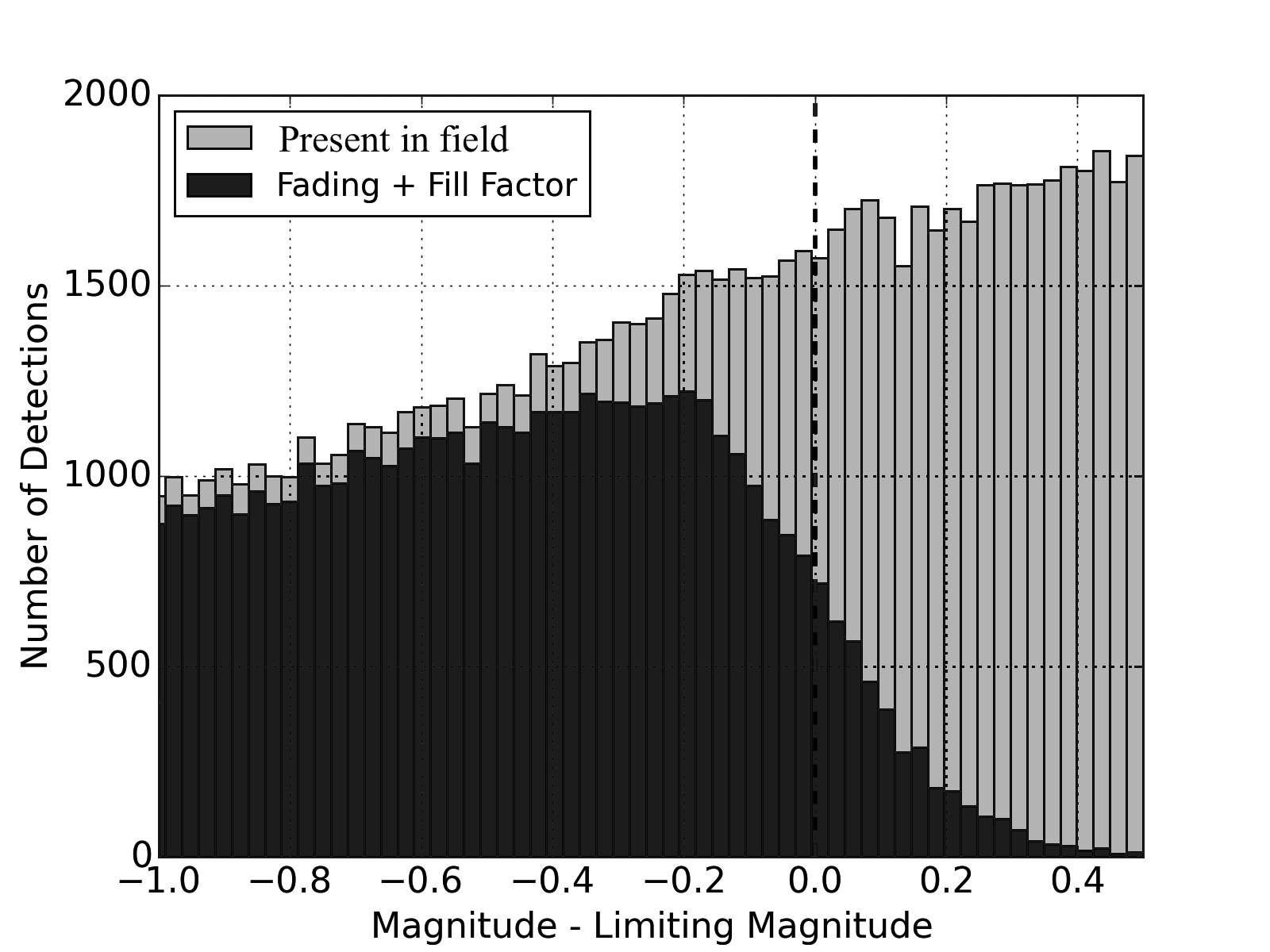}
     \caption{Histogram of NEOs present in the field in one night of the LSST survey (light bars) and detected NEOs with fading, $m_5$ and fill factor applied (dark bars). The cumulative number of NEOs detected with fading is slightly greater than with the $m_5$ step function scenario.}
    \label{fig.fading_gain}
\end{figure}

\subsection{Trailing and detection losses}

\begin{figure}[htb]
  \centering
    \plottwo{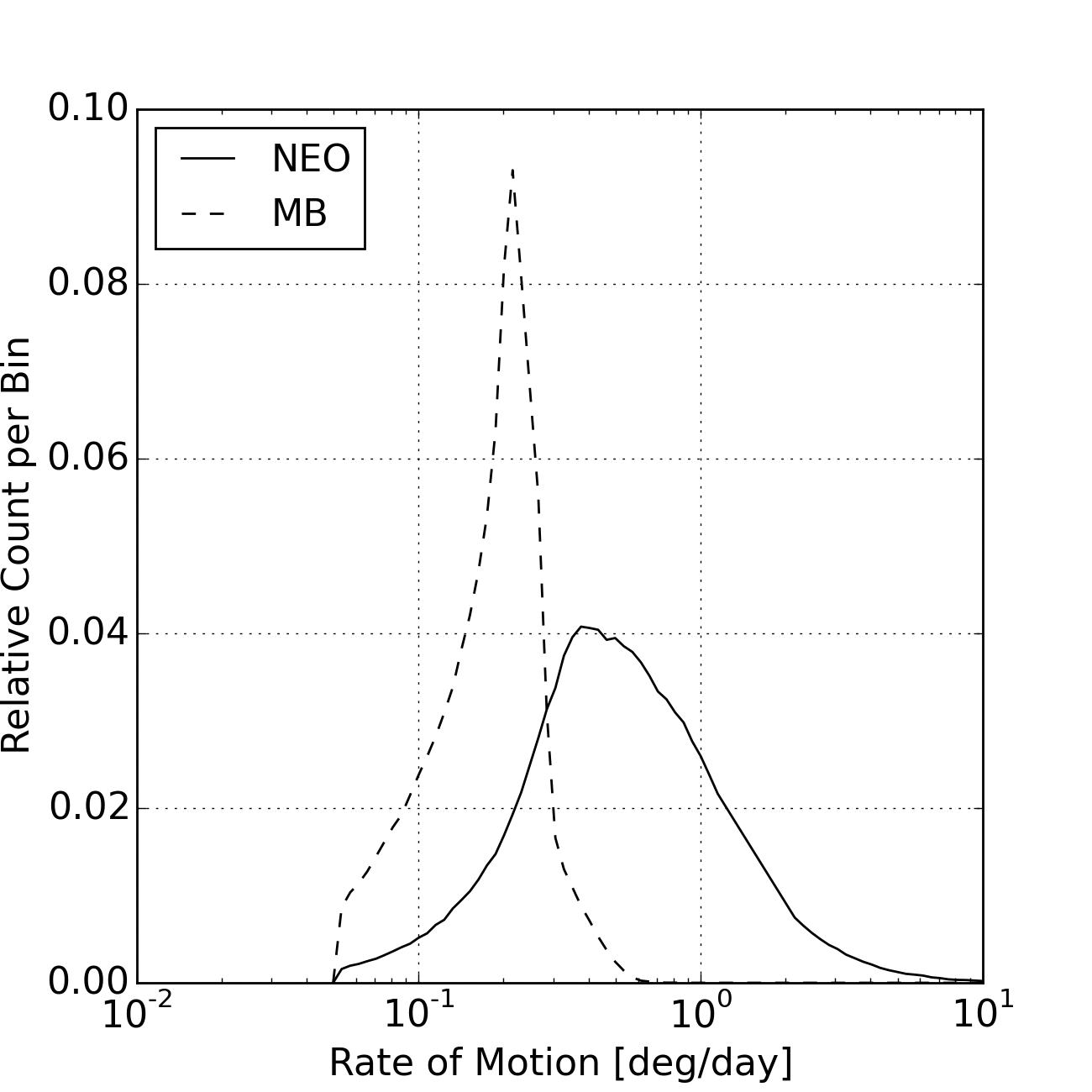}{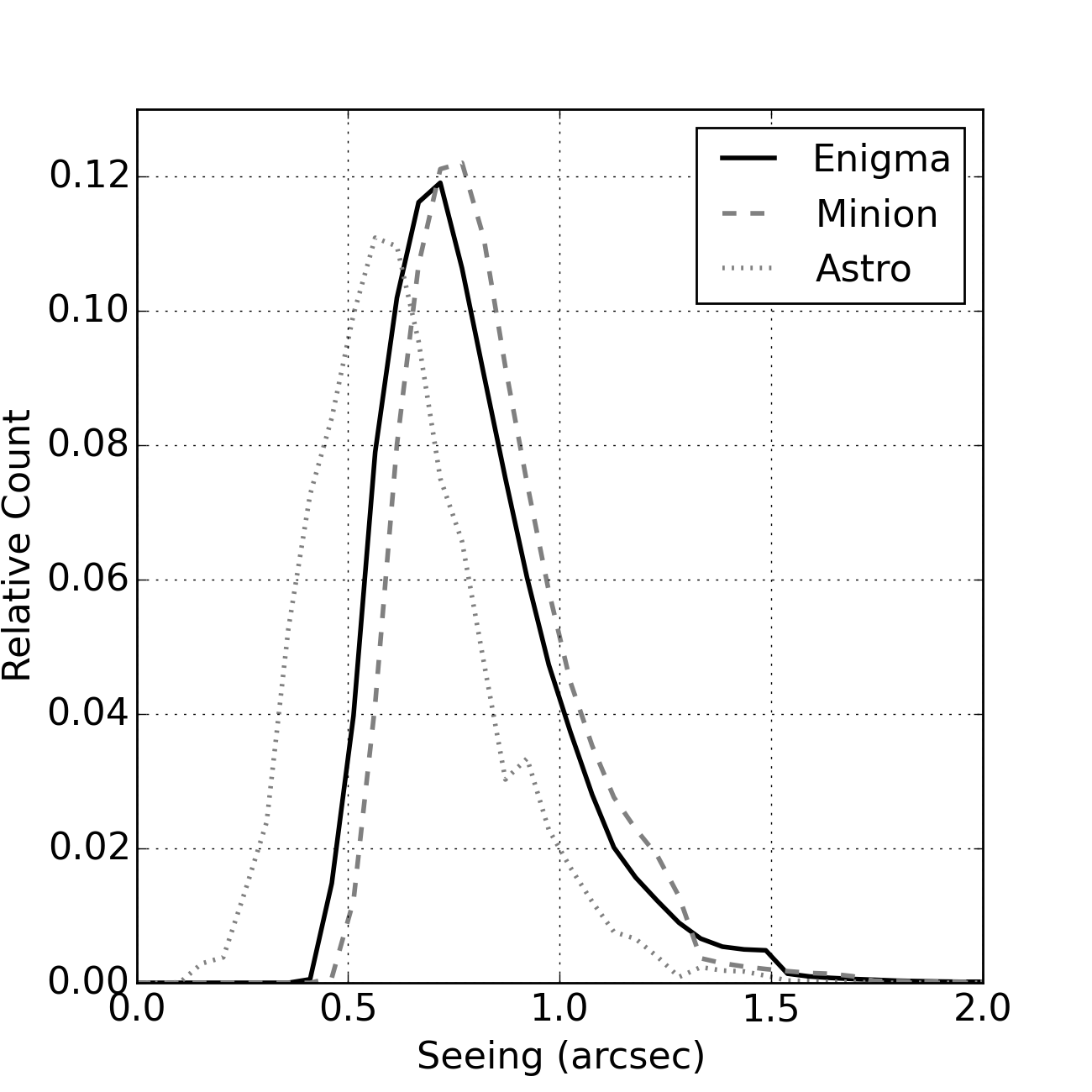}
     \caption{Left -Normalized histogram of rate of motion of main belt and Near-Earth asteroids detected by LSST in one lunation of \enig.  Right - Normalized histogram of seeing in 3 simulated surveys.}
    \label{fig.rates}
\end{figure}

\begin{figure}[htb]
  \centering
      \epsscale{0.5}
    \plotone{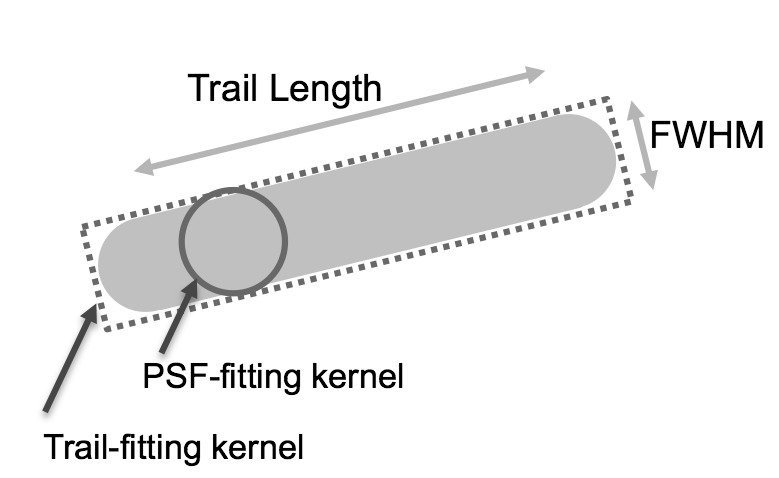}
     \caption{The profile of the trail and the PSF kernel.}
    \label{fig.trail}
\end{figure}

LSST will observe the sky with a sidereal tracking rate, and so the static sources will be detected as point-spread-functions (PSFs). Asteroids, however, will move in the images within the exposure time. For instance, in a 30-second single exposure a typical NEA moving at 0.64 deg/day (Figure~\ref{fig.rates}) will move by 0.8 arcsec. The expected mean seeing at the LSST site is 0.85 arcsec (Figure~\ref{fig.rates}), therefore,  instead of a PSF, most NEAs will appear trailed (Figure~\ref{fig.trail}), with the fastest objects being the smallest and closest to the Earth. The detected trail may be described as a convolution of a PSF and a line \citep{2012PASP..124.1197V}. The flux of the moving source is spread along multiple pixels and the per-pixel signal decreases as a function of its apparent velocity. The longer the trail, the fainter is the peak signal. Also, the signal-to-noise ratio (SNR) decreases due to trailing as the trailed detection contains the sum of noise from a greater number of pixels. 

We consider two types of trailing magnitude losses. The first one is the the loss that happens when a Gaussian or a PSF-like filter is used to identify the sources in the image. The source function is traditionally modeled according to the static and well defined sources like unsaturated stars \citep{jones2017}, therefore, if the model finds a trail, it only captures a fraction of the flux in it (Figure~\ref{fig.trail}). We call this magnitude loss the ``detection" loss \citep{2009AAS...21346003I} and it is described by the function
\begin{equation}
\label{eq.trailing_loss}
\Delta m_{Trail} = 1.25 \log_{10}{\left({1+cx^2}\right)}
\end{equation}
where $c=0.42$ and $x=(vt_{exp})/\Theta$ is the trail length in units of FWHM seeing disks $\Theta$. Here $v$ is the rate of motion and $t_{exp}$ the exposure time. 

\begin{figure}[htb]
  \centering
    \plotone{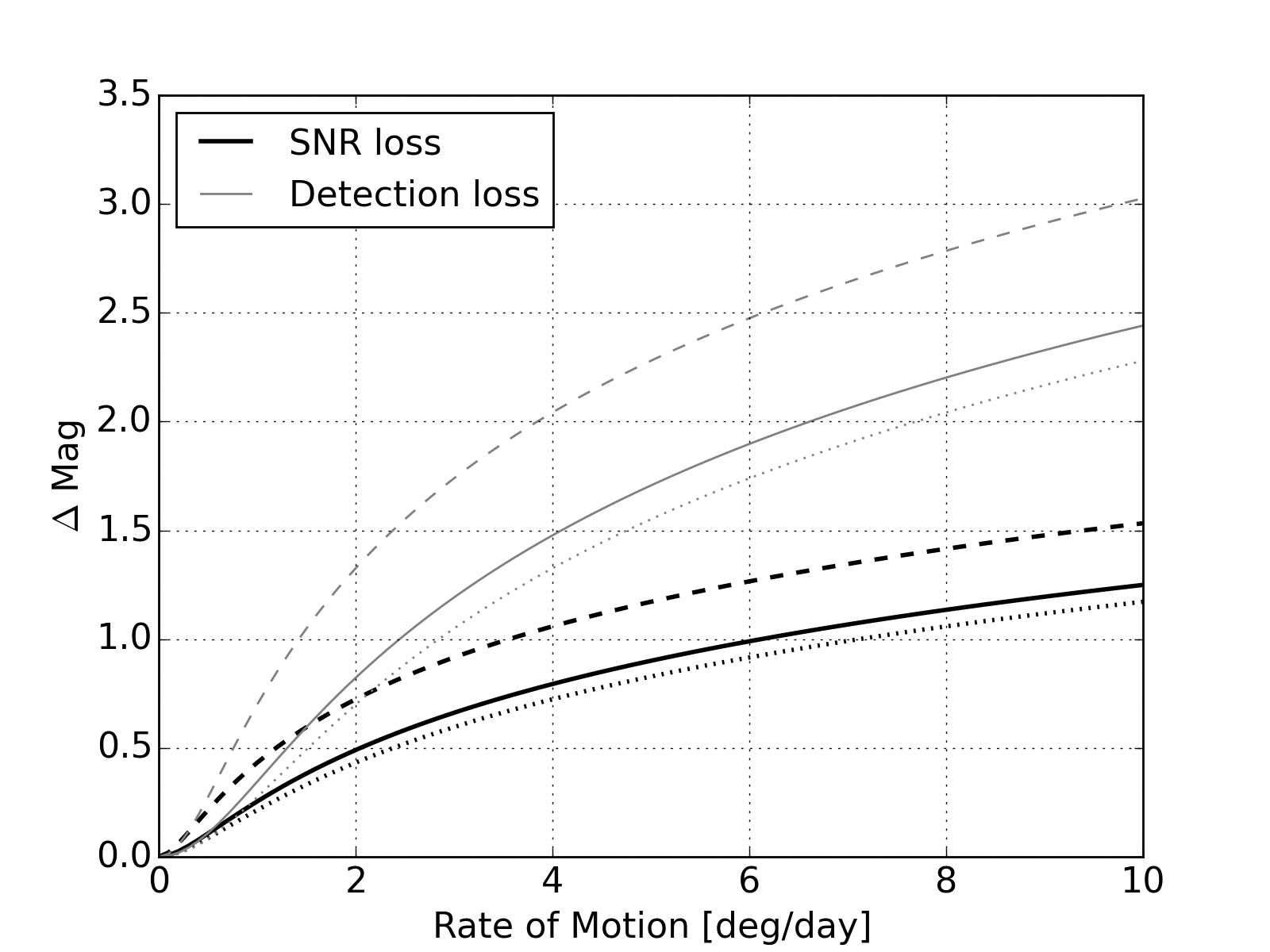}
     \caption{SNR and detection loss as a function of rate of motion for a 30-second exposure and average seeing of 0.86 arcsec (solid curves). The magnitude loss strongly depends on seeing. The dotted lines represent a substandard seeing (1.0 arcsec), dashed lines a superb seeing (0.6 arcsec). }
    \label{fig.trail1}
\end{figure}

After LSST detects a source, it calculates the SNR with a number of algorithms, including the use of a trail-fitting kernel, which leads to our second type of trailing magnitude loss. Here  the SNR of the trail is calculated from the source flux and the noise of the background from the entire trail. The greater the area of the source, the larger the amount of background noise that decreases the overall SNR of the detection. The magnitude loss due to the SNR trailing penalty \citep{2009AAS...21346003I} can be described by the function
\begin{equation}
\label{eq.snr_loss}
\Delta m_{SNR} = 1.25 \log_{10}{\left(\frac{1+ax^2}{1+bx}\right)}
\end{equation}
where $a=0.76$, $b=1.16$ and $x$ is again the normalized trailing factor \citep{jones2017}. As shown in Figure~\ref{fig.trail1}, the detection losses are a factor of two worse than the SNR losses, which immediately implies that significantly trailed detections found by LSST will have $\mathrm{SNR}\gg5$.

\newpage
\subsection{Asteroid Colors}

Absolute magnitudes of asteroids are published in Johnson V-band and are used to calculate the apparent magnitudes. However, LSST will observe in 5 distinct filters in the visible and near infrared (u,g,r,i,z,y) and the transformation to V-band depends on the spectral characteristics of the asteroid. Even though distributions of asteroid colors in the main belt are relatively well understood to $H<18$ \citep{2014Natur.505..629D}, sampling of the NEA population is rather sparse. For instance, the SDSS database \citep{2010A&A...510A..43C} contains only 174 NEA and the Asteroids Lightcurve Database \citep{2009Icar..202..134W} currently contains 1,115 NEA with identified spectral type. That is only a small fraction of the NEAs currently known (15,000 as of November 2016). Bottke's and Granvik's NEA populations, which we used in this work do not include albedo, diameter, or spectral type. We generated the distribution of spectral classes of NEA by the debiased distribution derived by \citet{2004Icar..170..295S} shown in Table~\ref{tab.NEO_classes}. For simplification of the magnitude transformation from LSST filters to Johnson V-band, we divide these classes into two groups: S (Q-type, S-type and high-albedo X-types) and C (C-types, D-types and low-albedo X-types). With this scheme, the numbers of C and S type asteroids are similar. Also, the generated colors do not depend on $H$ or orbital elements. The magnitude transformation model used in the study is presented in Table~\ref{tab.color}. The tabulated color indices are based on the LSST filter bandpasses, with mean reflectance spectra from \citet{demeo_pds_2009} and the \citet{chance_2010} model for the solar spectrum.
 
\begin{table}[H]
\small
\begin{center}
\caption{NEA Spectral Classes by \citet{2004Icar..170..295S}. Note, that X-class consist of two distinct subpopulations based on albedo.}
\begin{tabular}{c|l|c}
\tableline\tableline
 & \multicolumn{1}{c|}{Type} & Fraction\\
\tableline
\multirow{3}{*}{\rotatebox[origin=c]{90}{C Group}} & X \textsubscript{low albedo} & 0.17\\
& D & 0.18\\
& C& 0.10\\
\hline
\multirow{3}{*}{\rotatebox[origin=c]{90}{S Group}}&X \textsubscript{high albedo}&0.17\\
&Q &0.14\\
& S&0.23\\
\tableline
\end{tabular}
\label{tab.NEO_classes}
\end{center}
\end{table}

\begin{table}[H]
\small
\begin{center}
\caption{Magnitude transformation from Johnson's V-band to LSST filter system.}
\begin{tabular}{c|cccccc}
\tableline\tableline
 Asteroid color &V-u &V-g& V-r& V-i& V-z& V-y\\
\tableline
C Group & -1.614 & -0.302& 0.172& 0.291& 0.298& 0.303 \\
S Group & -1.927 & -0.395& 0.255& 0.455& 0.401& 0.406 \\
\tableline
\end{tabular}
\label{tab.color}
\end{center}
\end{table}

\subsection{Light Curve Variation}

The apparent magnitudes of asteroids derived from the MOPS ephemerides do not reflect the amplitude variation due to asteroid shapes, rotation and spin axis orientation. This effect could be significant when the sky is observed in a sparse temporal resolution, e.g., a few times per night. If the asteroid has an apparent brightness close to the limiting magnitude, a change in the brightness due to rotation in the time interval between two LSST exposures can cause the asteroid to be visible only in one of the images. Therefore, the tracklet will not be created and the asteroid would be lost on that particular night. On the other hand, some asteroids can be brighter than the ephemeris magnitude just because they were observed near the maximum of the light curve, leading to the possibility of finding objects nominally below the detection limit.

In this work we generated amplitudes and periods for the model populations based on the debiased model by \citet{2009Icar..204..145M}. We extended this model so that it depends on the absolute magnitude $H$ by the method described in \citet{2015Icar..261...34V}. The simplified light-curve corrections are represented by a sine wave defined by the generated amplitudes and periods at the epochs of the time of observation.  This approach does not reflect the real shape of asteroids nor an amplitude that depends on the phase angle or the spin axis orientation. Despite the simplification, the generated magnitudes should reflect reality because they follow the debiased not the observed distribution. The main drawback lies in the fact that \citet{2009Icar..204..145M} observed MBAs near opposition at very low phase angles, while NEA phase angles vary widely. At large phase angles and differing geometry, the resulting light curve is altered and the resulting amplitudes may be smaller.


\subsection{Vignetting}

\begin{figure}[H]
  \centering
  \epsscale{0.5}
    \plotone{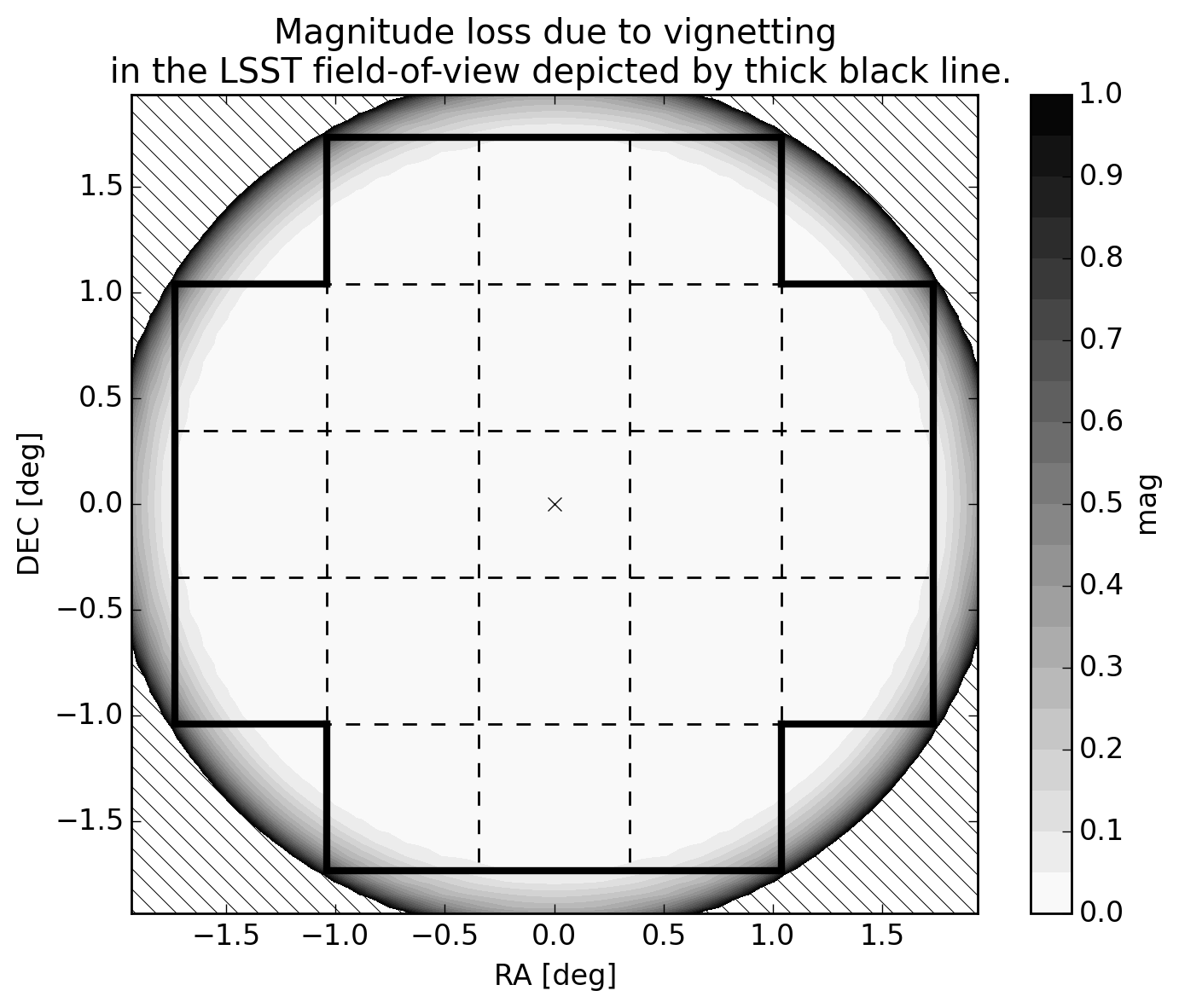}
     \caption{Contour plot showing magnitude loss due to vignetting \citep{2016SPIE.9906E..0LA} in the LSST focal plane, depicted by thick black line. Dashed lines represent individual rafts.}
    \label{fig.vignetting}
\end{figure}

Optical and mechanical pathways with lenses and mirrors cause vignetting, which  decreases brightness, especially at great distance from the optical axis of the system. Large focal planes and wide fields are especially prone to vignetting, even though specialized optical elements and optimized wide-field systems like LSST can reduce the effect significantly \citep{2015ApOpt..54.9045X}. The LSST vignetting model depends only on the distance from the center of the field \citep{2016SPIE.9906E..0LA}. Figure~\ref{fig.vignetting} shows the effect of vignetting causing magnitude loss in the LSST focal plane. Only 7\% of the collecting area has a sensitivity penalty due to vignetting that is greater than $0.1\,\mathrm{mag}$. The magnitude loss is significant only far from the field center and will only affect the most distant corners of the detector. Therefore, vignetting should not be expected to cause a large completeness loss.

\section{Simulations}

MOPS --- our simulation framework -- provides detected NEOs in LSST fields based on selected parameters and constraints. A list of detections for a given night is submitted to  MOPS in the second stage to create tracklets. A tracklet is generated for a detection in the first image of a given field if a second detection near the first is present in a subsequent image. The area of the search circle is determined by lower and upper velocity limits, set as 0.05 deg/day and 2.0 deg/day in this work. If there are more possible connections in the circle, in addition to a clean tracklet consisting of the same object,  a tracklet consisting of two objects is created as well. Increasing the upper velocity limit increases the number of false tracklets rapidly. Therefore, for velocities of 1.2--2.0 deg/day we used the information on the trail length of the detection for making tracklets. At 1.2 deg/day, the detection will have a non-PSF shape and its length will be 1.8 times the PSF at  the average 0.86 arcsec seeing, and so its length and orientation can be determined. Thus, instead of a large circular search area around detections that are trailed, only smaller regions, consistent with the anticipated velocity and direction of the trails need to be searched. Potential matches must also agree in the rate of motion.

\begin{figure}[H]
  \centering
      \epsscale{0.5}
    \plotone{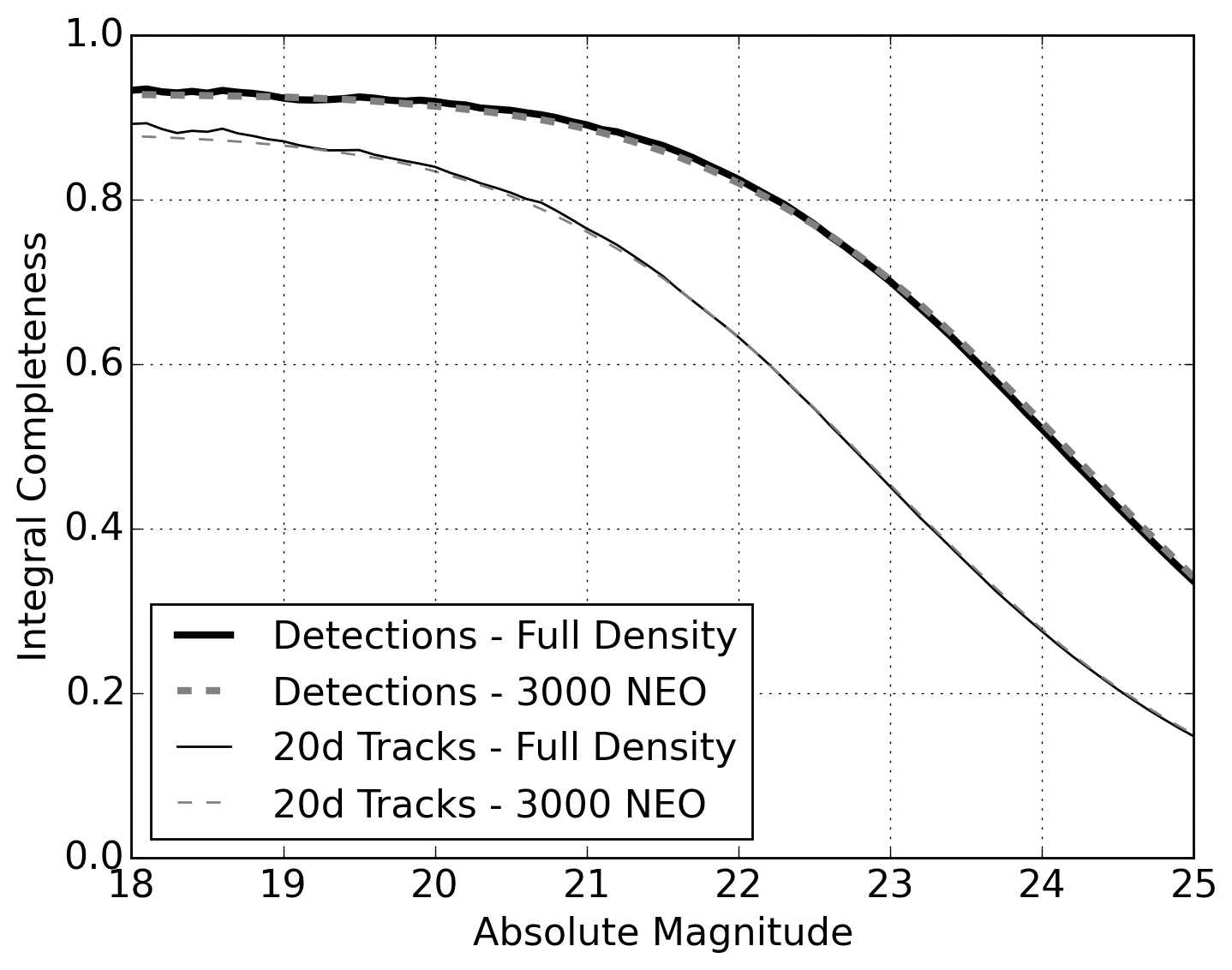}
     \caption{Comparison of completeness for a full-density NEO model (solid) and a 3000 object random sample (dashed). Upper curves represent objects detected in at least one detection in 10 years and the lower curves require a track of 3 tracklets within 20 days to be present. Assumed 92\% fill factor, with fading and trailing losses.}
    \label{fig.low_vs_high_fidelity}
\end{figure}

Having detection and tracklet lists is enough to predict tracks, neglecting linkage inefficiencies. Moreover, to perform quick studies of the 10-year survey detection and tracklet efficiency, we decreased the number of NEO orbits to 3000 and created detection and tracklet lists with only limiting magnitude and field shape and rotation parameters. The absolute magnitude for all 3000 orbits was set to $H=0$. The advantage of this approach is that allows us to adjust individual model parameters in post-processing and measure their effects on overall performance. The per-bin detection efficiency is then derived from the list of simulated detections or tracklets by adding a bin-by-bin $\delta H$ correction according to the anticipated size-frequency distribution and then accumulating the found objects to obtain integral completeness. This low-density approach is $>100\times$ faster in comparison with a full-density NEO-only simulation. Ephemerides of 3000 representative NEO orbits were computed for all OpSim fields, providing a detection list. Post-processing readily creates tracks because all detections and tracklets are identified with the associated object by MOPS. We studied two scenarios for potential tracks: 12-day and 20-day tracks. The minimum number of distinct nights for a track is 3, and so the minimum time span between the first and last tracklet is about 2 days. The output of this quick sim is an integral completeness that agrees well with the full density sim (Figure~\ref{fig.low_vs_high_fidelity}), and so we rely primarily on the low-density approach in this paper. The output of the low-density simulation is an integral completeness for detections, tracklets and tracks as a function of absolute magnitude or time (Figure~\ref{fig.compare}).

\subsection{System Performance and Model Dependencies}
\begin{figure}[p]
  \centering
\includegraphics[width=0.75\textwidth]{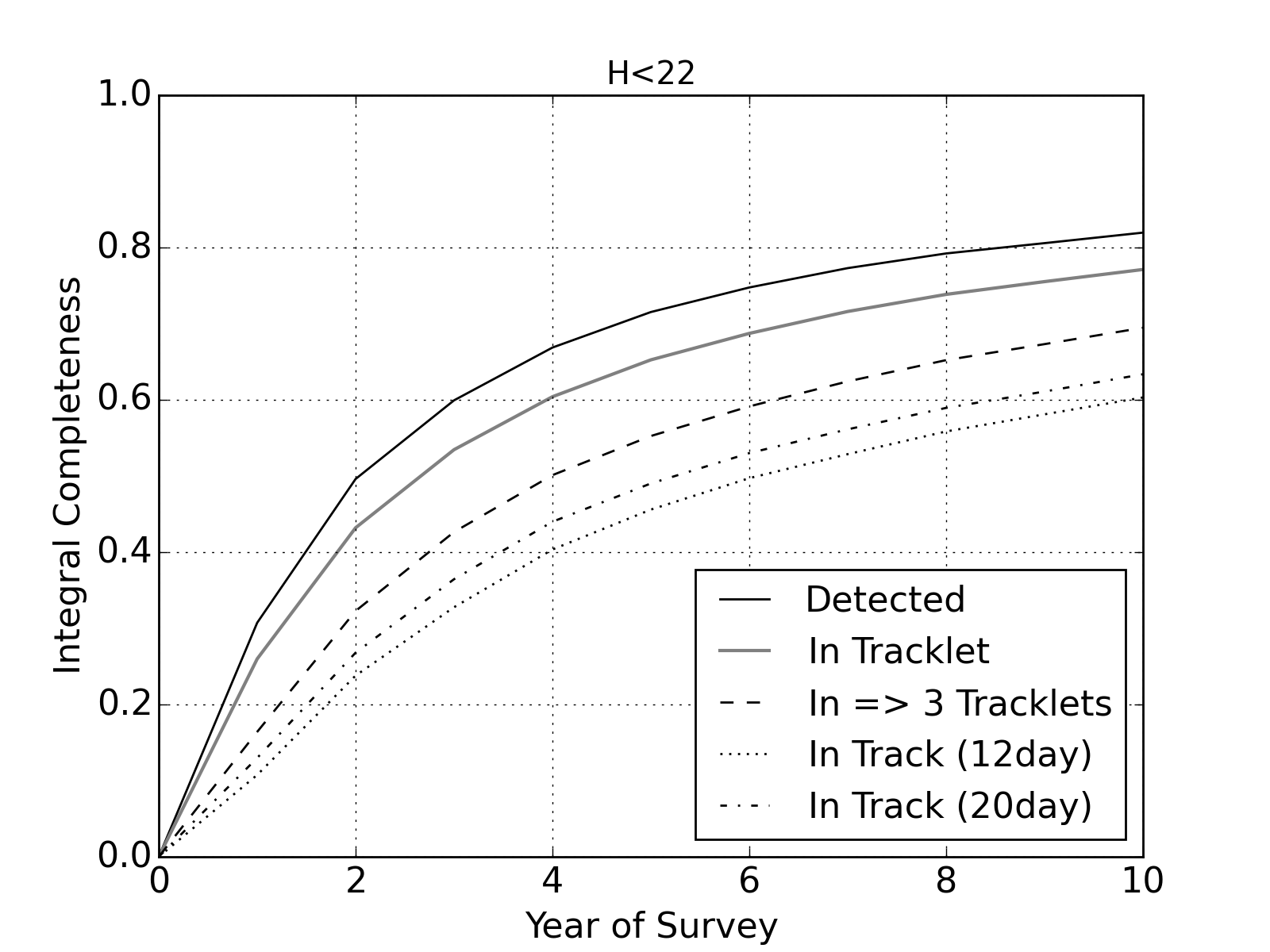}
\includegraphics[width=0.75\textwidth]{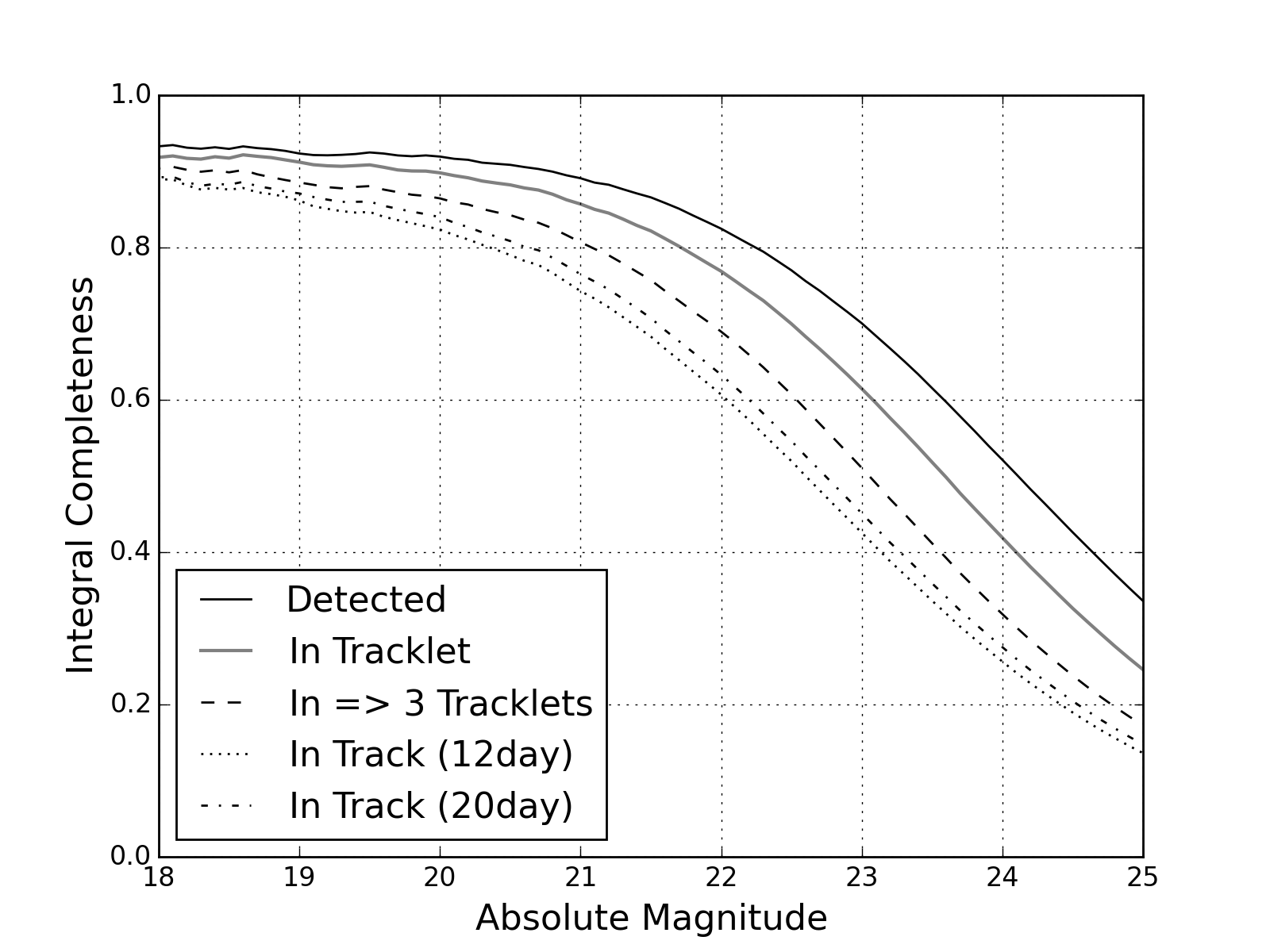}
         \caption{Detection, tracklet and track efficiency (NEO completeness) in a low-density simulation as a function of time for $H<22$ (top) and absolute magnitude after 10 years (bottom) in the \enig baseline survey. }
    \label{fig.compare}
\end{figure}

The key output of our simulation is the 10-year integral completeness for detections, tracklets and tracks as a function of absolute magnitude or time. Figure~\ref{fig.compare} reveals that over 80\% of NEOs with $H<22$ are detected at least once in the ten-year \enig survey. That number drops to 76\% for at least one tracklet in ten years, and 67\% for three tracklets in ten years, irrespective of their timing. To actually consider an object discovered and cataloged in this study, we require three tracklets on distinct nights over no more than 20 days or 12 days, which leads to 61\% or 58\% completeness, respectively, in Figure~\ref{fig.compare}. The time history of completeness reveals that after ten years the rate of cataloging is still increasing at about 2\% per year in \chtt.

We tested the importance to \chtt of a number of our modeling assumptions as discussed in the following list. The relevance of each item for \chtt is tabulated in Table~\ref{tab.NEOmodels} and depicted in Figures~\ref{fig.pop_FF_trailing_fading} and \ref{fig.color_light_vig_limmag}

\begin{table}[p]
\small
\begin{center}
\caption{\chtt completeness sensitivity due to individual parameters for \enig survey.}
\begin{tabular}{c|c|ccccc}
\tableline\tableline
Parameter& Penalty& Detection &Tracklet & 3 Tracklets & 12-day track& 20-day track\\[-3mm]

& & \%& \%& \%& \%& \%\\
\tableline
\tableline
\multirow{2}{*}{Population penalty} & Bottke 's NEOs&  0 & 0 & 0 & 0 & 0 \\[-3mm]
& Granvik's NEOs & $+0.7$&$+1.0$ &$+1.4$ &$+1.7$ &$+1.6$  \\
\hline
\multirow{3}{*}{Fill Factor} & 0.90 & 0 & 0 & 0 & 0 & 0 \\[-3mm]
&0.89& $0.0$&$-0.1$ &$-0.2$ &$-0.5$ &$-0.4$  \\[-3mm]
& 0.88  & $0.0$&$-0.1$ &$-0.3$ &$-0.9$ &$-0.7$  \\
\hline
\multirow{3}{*}{Trailing losses} & Off  & 0 & 0 & 0 & 0 & 0 \\[-3mm]
 &SNR loss & $-1.0$&$-1.3$ &$-1.6$ &$-1.8$ &$-1.8$  \\[-3mm]
&Detection loss & $-1.3$&$-1.6$ &$-1.8$ &$-2.1$ &$-2.0$  \\
\hline
\multirow{2}{*}{Fading}& Off & 0 & 0 & 0 & 0 & 0 \\[-3mm]
& On &$+0.9$&$-0.2$ &$-0.4$ &$-0.8$ &$-0.6$  \\
\hline
\multirow{3}{*}{Colors} & S class & 0 & 0 & 0 & 0 & 0 \\[-3mm]
&C+S class  & $-0.3$&$-0.3$ &$-0.4$ &$-0.5$ &$-0.5$\\[-3mm]
 &C class  & $-0.6$&$-0.7$ &$-0.9$ &$-1.2$ &$-1.2$\\
 \hline
\multirow{2}{*}{Light curve} & Off & 0 & 0 & 0 & 0 & 0 \\ [-3mm]
 &On & $+0.3$&$+0.1$ &$-0.1$ &$-0.3$ &$-0.2$  \\
\hline
\multirow{2}{*}{Vignetting} & Off   & 0 & 0 & 0 & 0 & 0 \\[-3mm]
 &On &$-0.2$&$-0.2$ &$-0.3$ &$-0.3$ &$-0.3$  \\
\hline
\multirow{3}{*}{Limiting magnitude} &$m_5$& 0 & 0 & 0 & 0 & 0 \\ [-3mm]
& $m_5-0.2\,\mathrm{mag}$ & $-1.7$&$-2.1$ &$-2.7$ &$-3.2$ &$-3.1$  \\[-3mm]
& $m_5 - 0.5\,\mathrm{mag}$ & $-4.5$&$-5.7$ &$-7.3$ &$-8.2$ &$-7.9$  \\
\hline
\multirow{4}{*}{Rate cut-off} & 0.5 deg/day& $-9.5$&$-9.6$ &$-12.3$ &$-13.6$ &$-13.3$  \\[-3mm]
&1.0 deg/day& $-2.0$&$-1.1$ &$-1.6$ &$-1.9$ &$-1.7$  \\[-3mm]
& 2.0 deg/day& 0 & 0 & 0 & 0 & 0 \\[-3mm]
&5.0 deg/day& $+0.5$&$+1.1$ &$+1.6$ &$+0.9$ &$+1.1$  \\
\tableline
\end{tabular}
\label{tab.NEOmodels}
\end{center}
\end{table}

\begin{itemize}

\item To compare the population models, we performed two low-density simulations with 3000 NEOs on \enig fields for Bottke's and Granvik's NEO models. Granvik's population led to a slightly greater completeness \chtt, but the efficiency is significantly greater for Bottke's population at $H<25$ (Figure~\ref{fig.pop_FF_trailing_fading}), primarily because most of Granvik's NEO are small and therefore much harder to be detected. Table~\ref{tab.NEOmodels} shows the percentage difference between the two models for detections, tracklets and tracks. 

\item Fill factor is one of the key effects that drives detection efficiency down. The low-density simulation focused on altering statistical fill factor by one percentage point.  Dropping fill factor by one and two percentage points from 0.90 led to almost no loss for detections and small losses (0.5-0.9\%) for 12- and 20-day tracks. (Figure~\ref{fig.pop_FF_trailing_fading}, Table~\ref{tab.NEOmodels})

\item Trailing losses represent a major effect for NEOs; however, are negligible for distant asteroids like MBAs. NEO completeness is reduced by 1 percentage point for single detections and up to 2 percentage points for 12- and 20-day tracks. Even though detection losses cause much larger magnitude loss per a single detection, in a 10-year low-density simulation, detection losses are similar to SNR losses in completeness. (Figure~\ref{fig.pop_FF_trailing_fading}, Table~\ref{tab.NEOmodels})

\item Because of Malmquist bias, the fading model leads to more detections than a hard step-function faint limit. However, for tracks, fading actually decreased the completeness. This nonintuitive effect arises because at the faint limit the fading model behaves just as a very low (50\%) fill factor. So, even though fading provided more detections, these originated from the faint end, below the $m_5$ threshold, where the probability of a single detection is below 0.5. In case of a tracklet, this is less than  $0.5^2$ and in case of tracks, that require 3 tracklets and minimum of 6 detections, the probability is only $0.5^6$ of getting cataloged. (Figure~\ref{fig.pop_FF_trailing_fading}, Table~\ref{tab.NEOmodels})

\begin{figure}[tb] 
  \centering
    \includegraphics[width=0.47\textwidth]{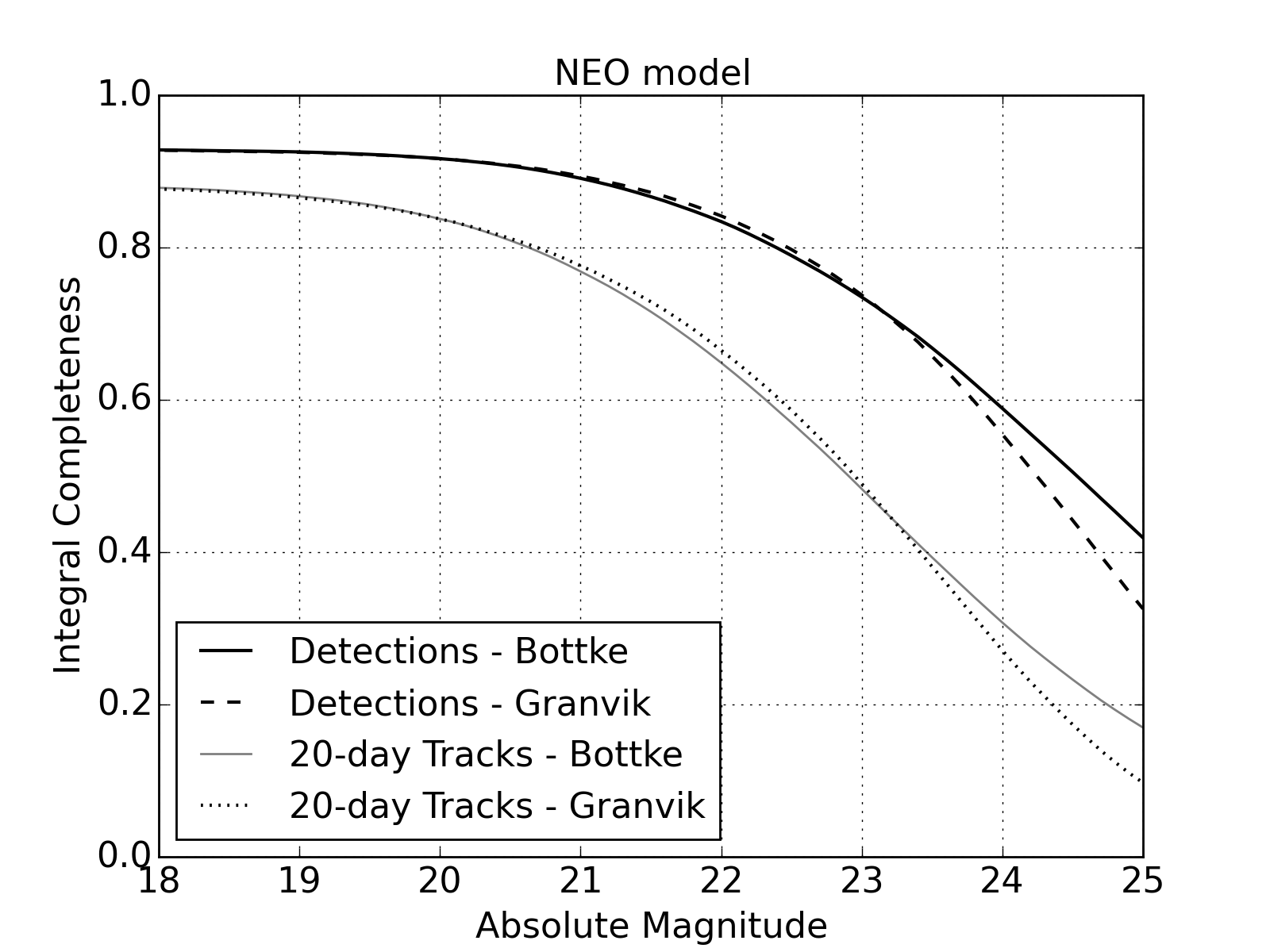}
    \includegraphics[width=0.47\textwidth]{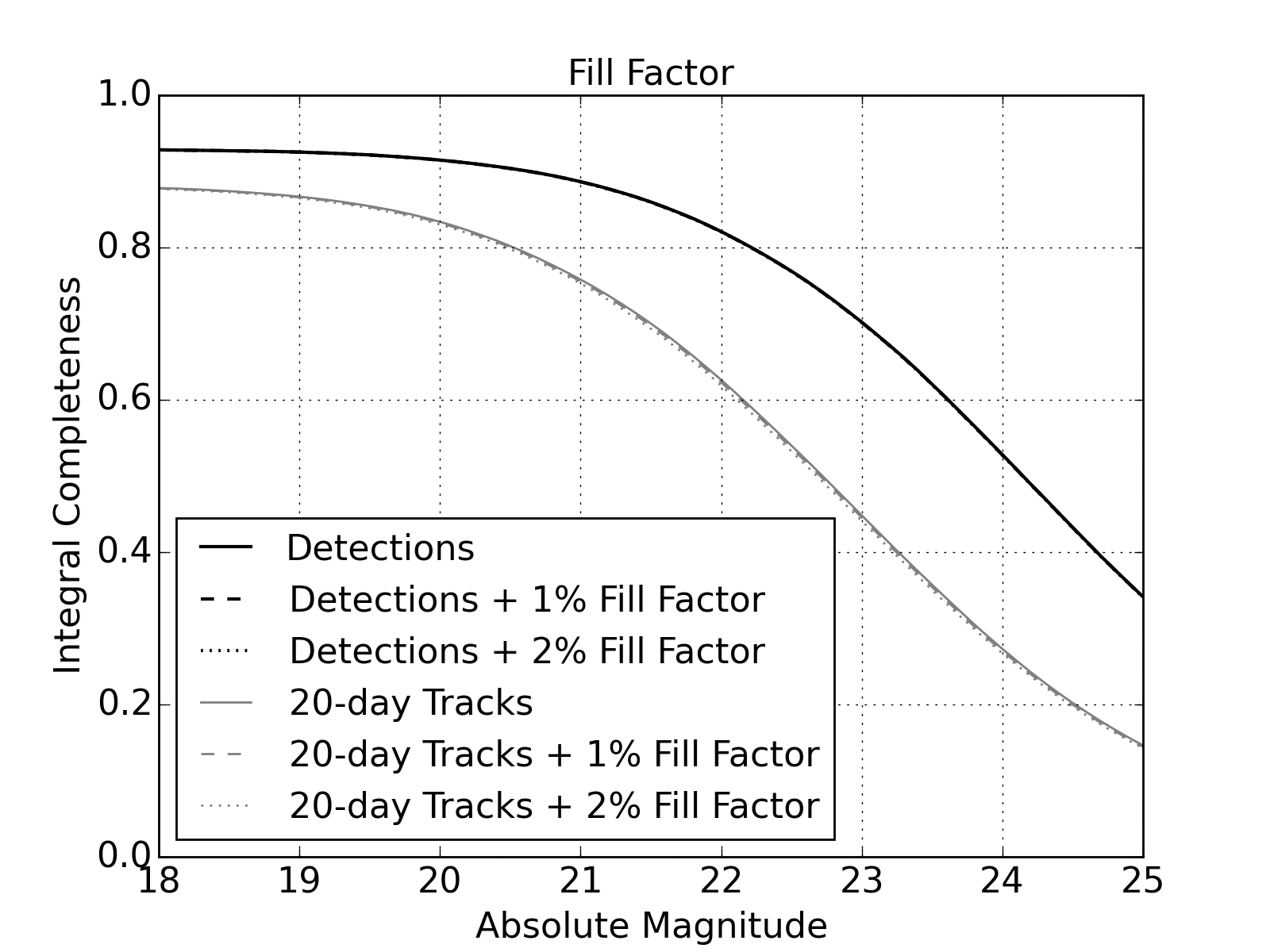}
    \includegraphics[width=0.47\textwidth]{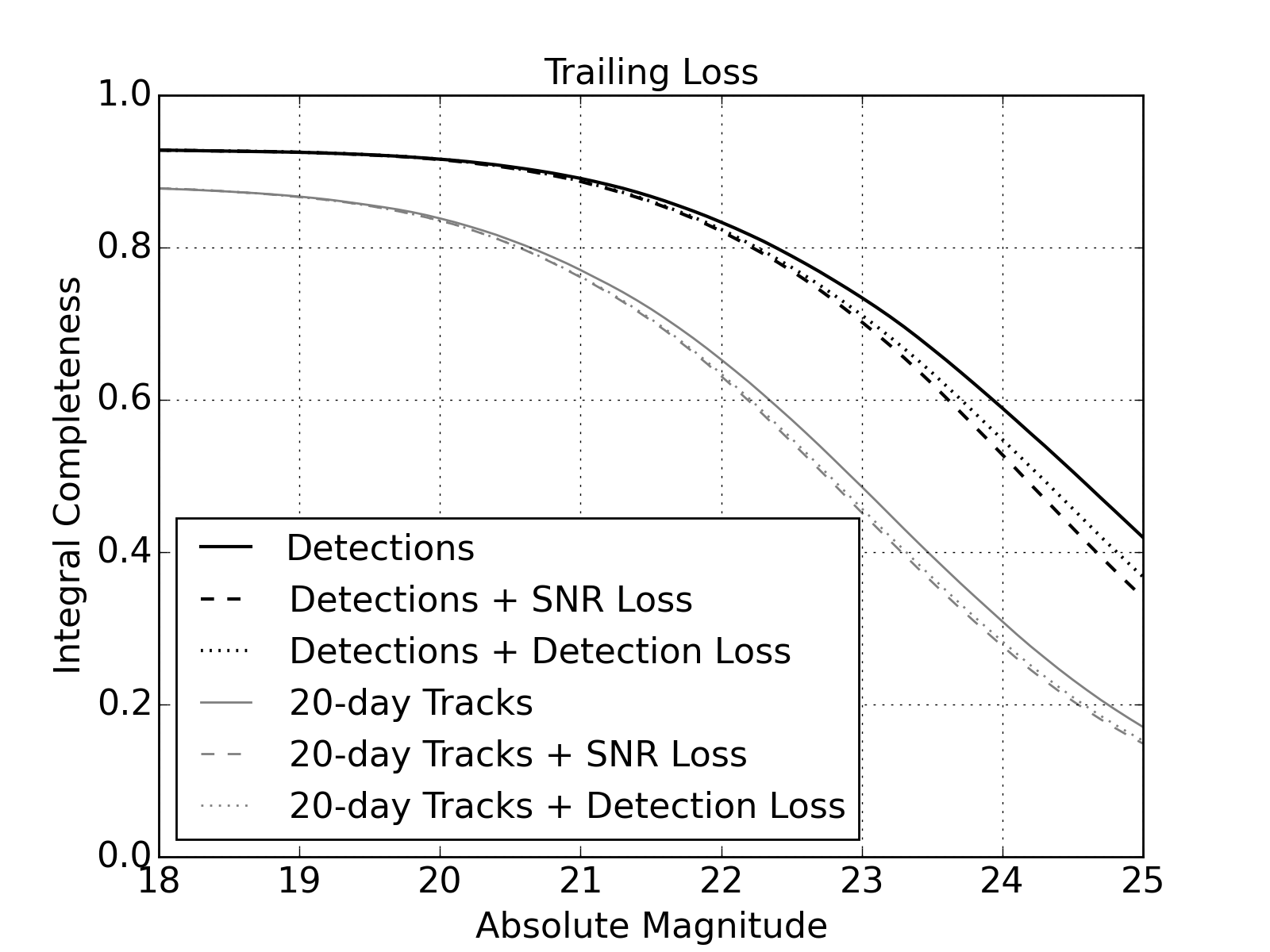}
    \includegraphics[width=0.47\textwidth]{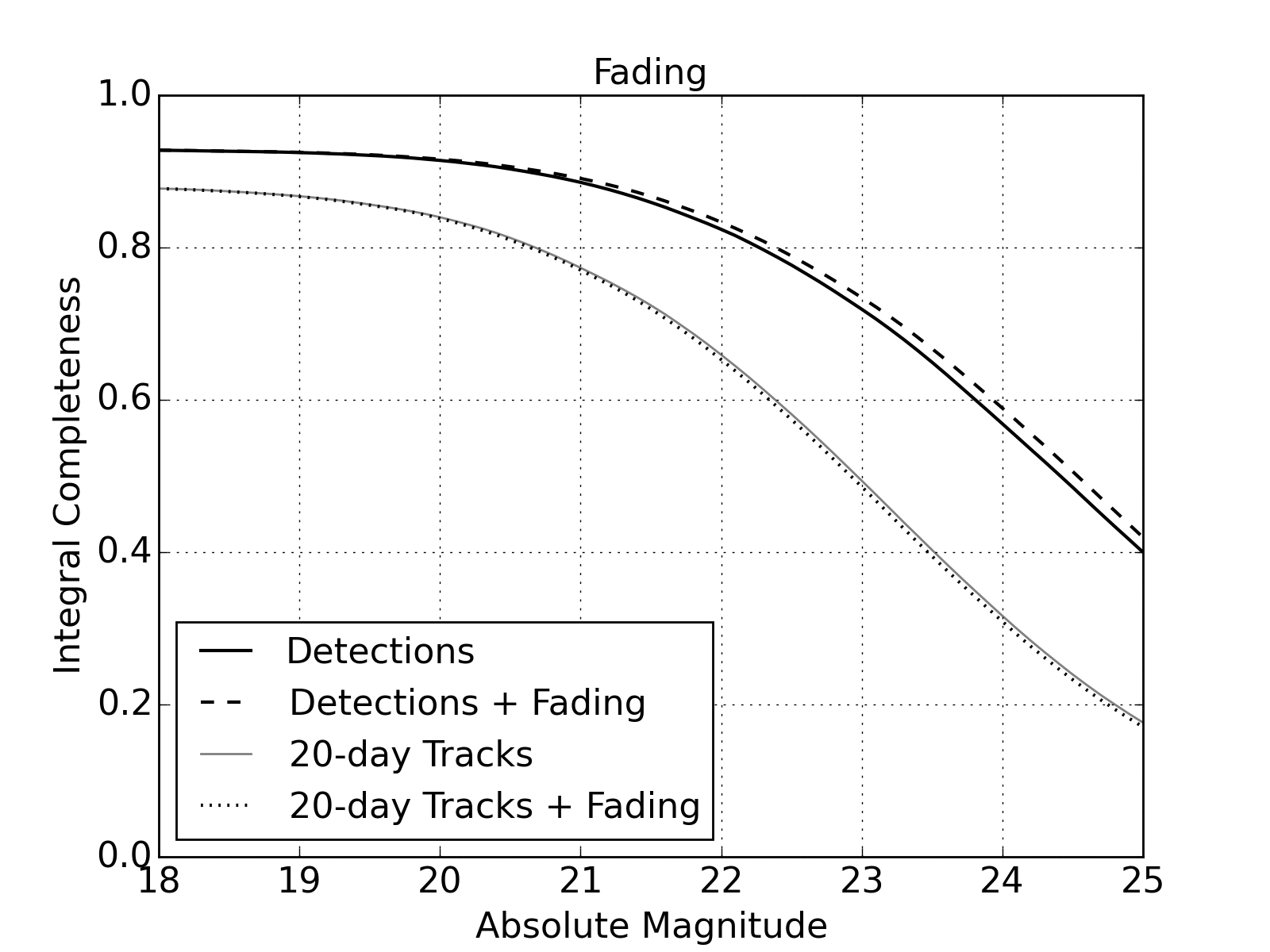}
     \caption{Effect of using two different NEO populations, effects of fill factor, trailing losses and fading on detections and tracks in a low-density simulation. Fill factor 0.90, \enig with Bottke's NEOs.}
    \label{fig.pop_FF_trailing_fading}
\end{figure}

\item We tested three scenarios with NEOs being only S type, C type or a 50-50 mixture of S and C types, independent of orbital elements or $H$. The default class of NEO used in this work is S types. Switching to C types, led to a net decrease of detection efficiency. Nevertheless, the loss was rather small. Switching from S to mixture of C and S types led to a relative loss of 0.5\% at $H<22$ for 2-day tracks. If all asteroids were C types, then the completeness loss will be slightly larger, at 1.2\%.(Figure~\ref{fig.color_light_vig_limmag}, Table~\ref{tab.NEOmodels})

\item Like fading, the light curve variation provided somewhat more detections and yet fewer linkable tracks. For detections and tracklets, the completeness increased by 0.3\% and 0.1\% at $H<22$. However, 12- and 20-day tracks showed that the light curve variability caused a decrease in completeness by 0.3\% and 0.2\%. This effect is similar to the detection gain due to fading and related combinatorics. The results showed that light curve variation has a negligible effect in completeness. (Figure~\ref{fig.color_light_vig_limmag}, Table~\ref{tab.NEOmodels})

\item As expected, vignetting plays only a minor role in the completeness of the survey. The NEO completeness penalty at $H<22$ is only about 0.3\% for the tracks. (Figure~\ref{fig.color_light_vig_limmag}, Table~\ref{tab.NEOmodels})

\item So far we have discussed several nominal modeling aspects of the baseline LSST survey. But we wish to also consider off-nominal performance of the LSST system. For instance, the limiting magnitude and seeing for fields is theoretical, even though based on long term observations from Cerro Pach\'{o}n. Systematic offsets in limiting magnitude could cause a drastic drop in completeness, possibly more than all previously mentioned parameters. For instance, if the limiting magnitude is only 0.2 mag shallower, the NEO ($H<22$) completeness penalty is $\sim 3\%$ for 12 and 20-day tracks, and for a 0.5 mag loss the corresponding penalty is $\sim 8\%$. (Figure~\ref{fig.color_light_vig_limmag}, Table~\ref{tab.NEOmodels})

\item This study enforced a tracklet velocity limit of 2.0 deg/day and a subset of that velocity range (1.2--2.0 deg/day) used a special treatment when creating tracklets, using the length and orientation of the elongated detections. Figure~\ref{fig.rates} depicts the rates of motion of NEOs and MBAs. Linking complexity is reduced for lower limiting velocities because pairs of detections are more separated at higher velocities, but this comes at the cost of reduced completeness because some fast moving asteroids would not be discovered due to the velocity limit. So selecting the velocity threshold is a tradeoff between computational loads and the objective of discovering small, fast-moving NEOs.Figure~\ref{fig.FF_rates} and Table~\ref{tab.NEOmodels} show how different velocity cutoffs for tracklets affect the NEO completeness. Decreasing the upper velocity limit to 0.5 deg/day would dramatically decrease \chtt, by more than 13\% for tracks. On the other hand, increasing the upper bound from 2.0 to 5.0 deg/day only had a slight benefit of about 1\% in completeness for tracks. If making use of the trail velocity information for fast detections, increasing the upper velocity threshold is not likely to significantly increase the false tracklet rate, and it comes at a modest benefit.

\end{itemize}

\begin{figure}[tb] 
  \centering
    \includegraphics[width=0.47\textwidth]{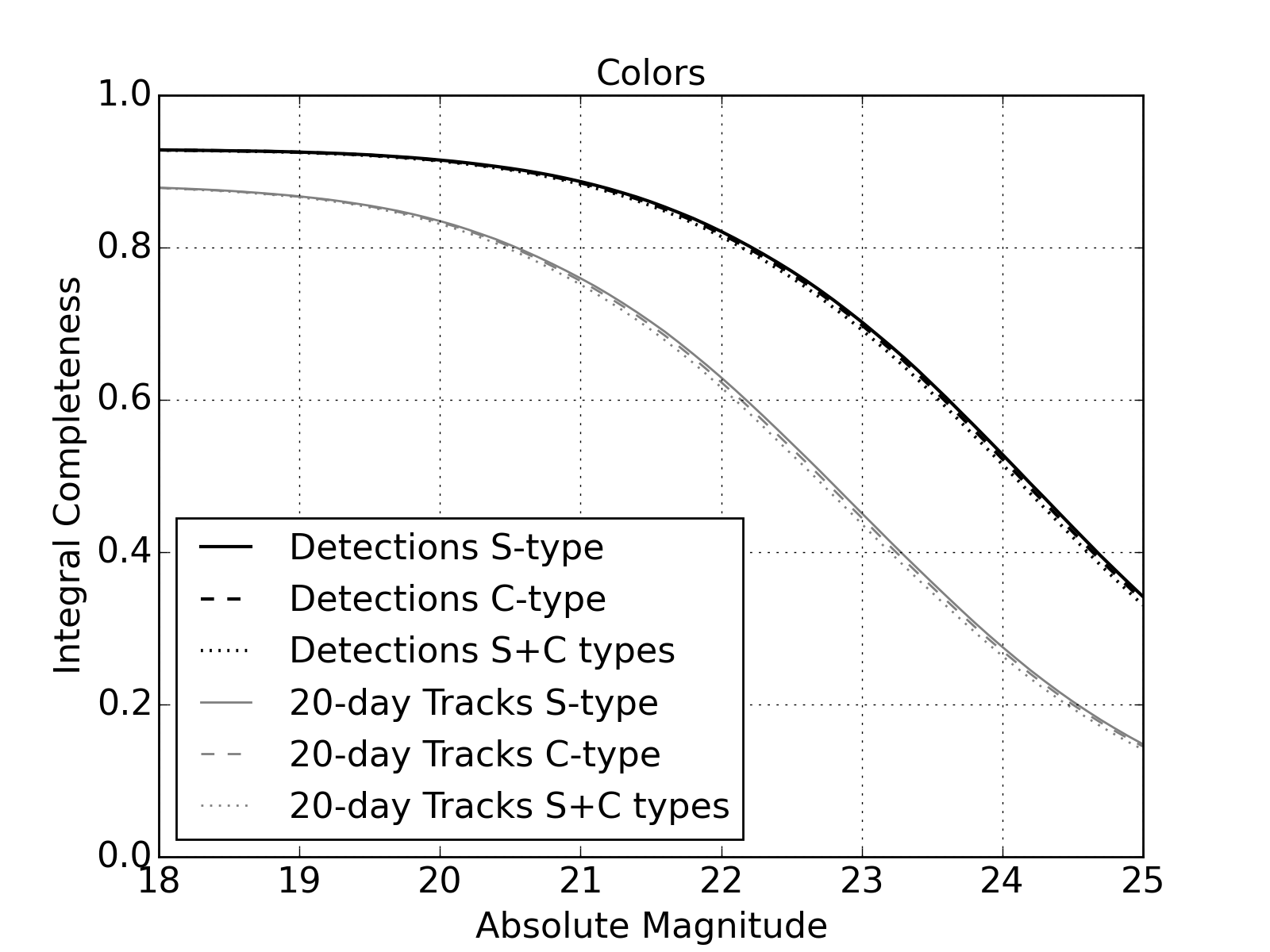}
    \includegraphics[width=0.47\textwidth]{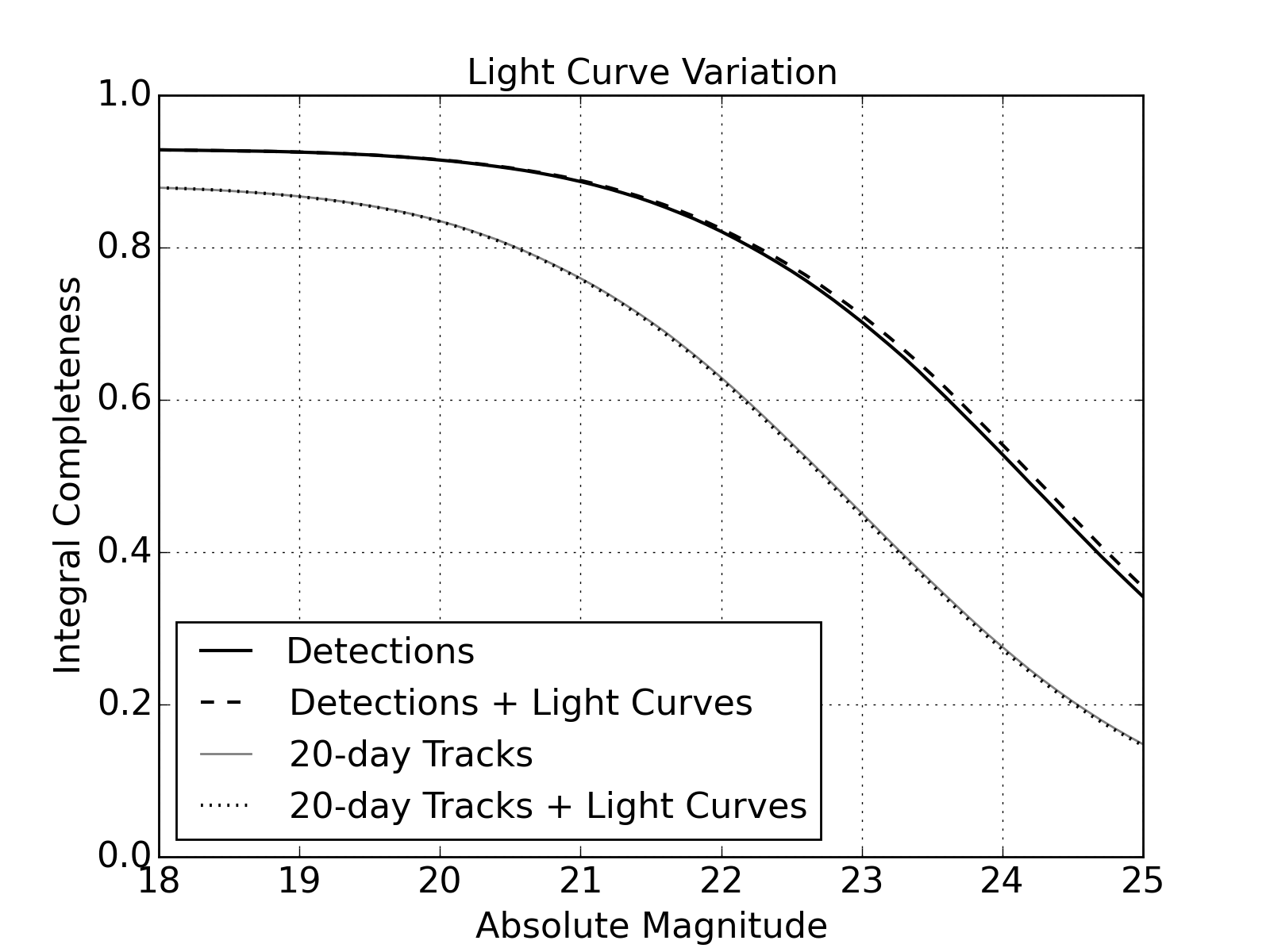}
    \includegraphics[width=0.47\textwidth]{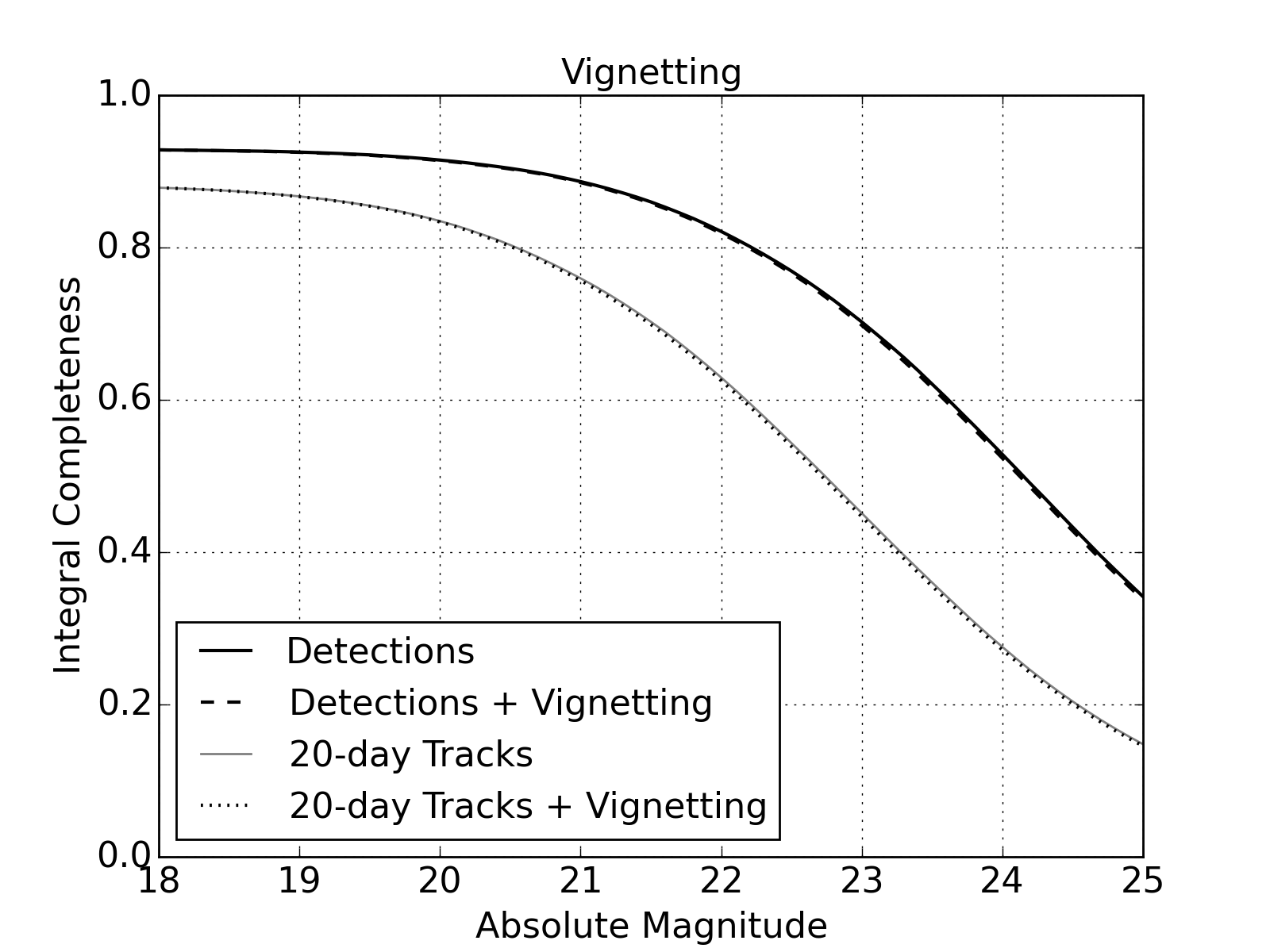}
    \includegraphics[width=0.47\textwidth]{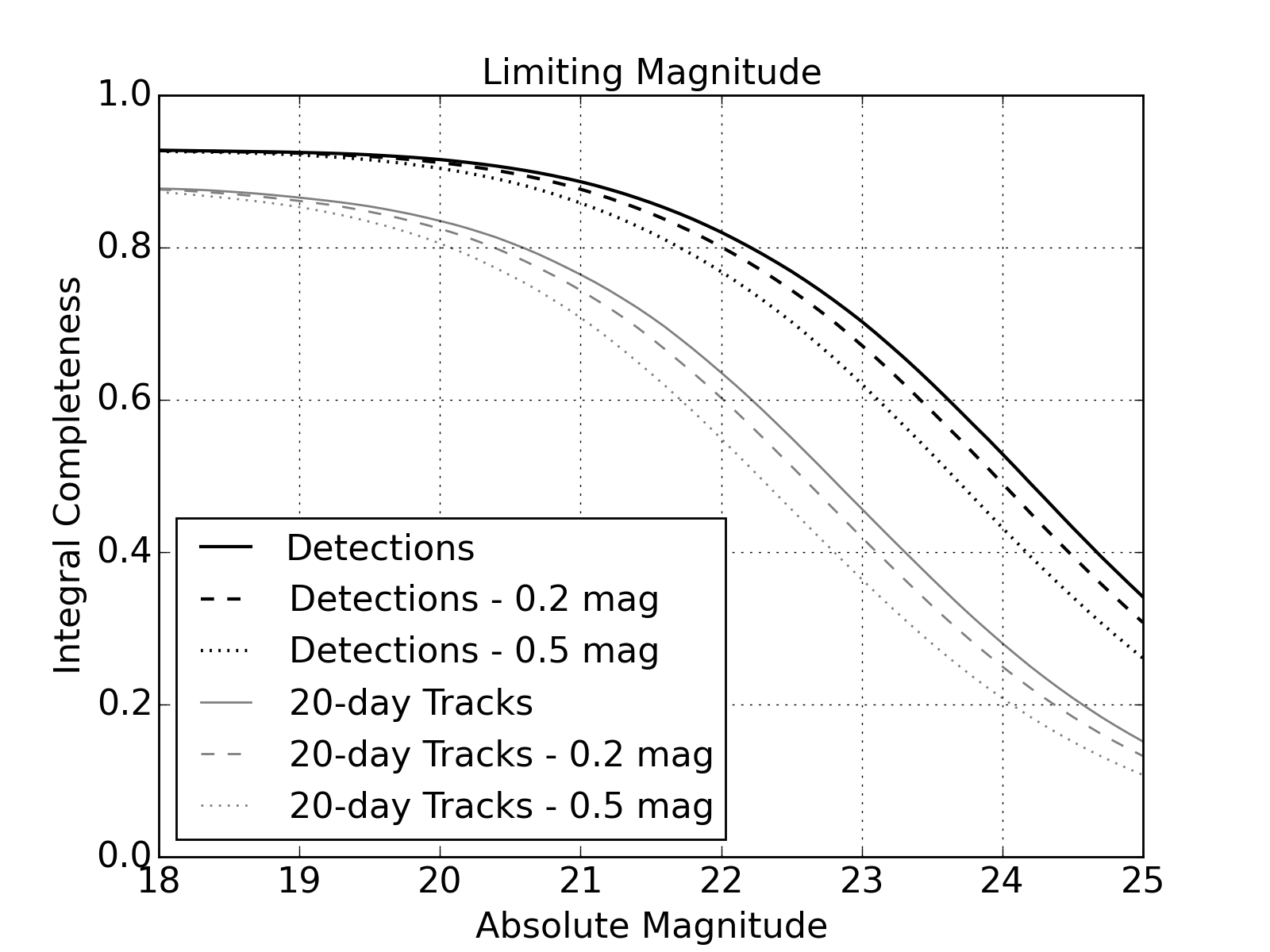}
     \caption{Effect of asteroid colors, light curve variation, vignetting and systematic penalty in limiting magnitude in a low-density simulation. Fill factor 0.90, \enig with Bottke's NEOs.}
    \label{fig.color_light_vig_limmag}
\end{figure}

\begin{figure}[tb]
   \centering
    \includegraphics[width=0.6\textwidth]{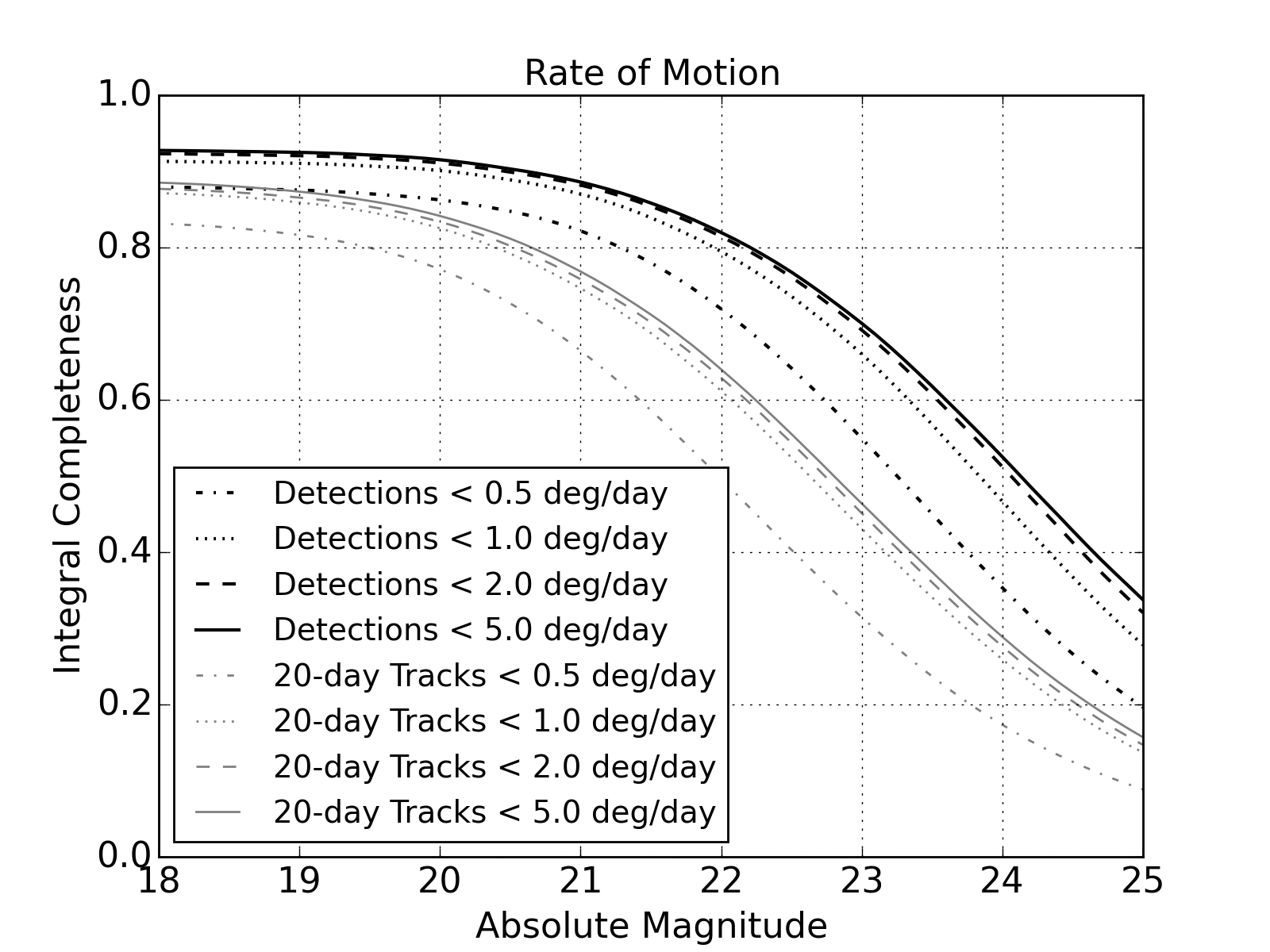}
     \caption{Effect of velocity thresholds on detections and tracks in a low-density simulation. Fill factor=0.90, for \enig, with vignetting, light curves, C+S colors, detection losses and Bottke's NEO.}
    \label{fig.FF_rates}
\end{figure}

\newpage
\subsection{Overall LSST Performance and Uncertainties}


Throughout this work we have tested and analyzed an array of different survey models, including various OpSim runs, NEO population models and detection models. Here we collect and summarize the various modeling details discussed above and characterize the uncertainty of the overall performance estimate.

For what we consider our final result we select the most current LSST baseline survey, \minion, and the latest debiased NEO population estimate, namely Granvik's NEO model. This ``new'' approach can be compared with the ``original'' alternative of \enig with Bottke's NEO model. Table~\ref{tab.enigma_vs_minion} makes this comparison in an incremental fashion by cumulatively adding various model details. Examining the \enig column of Table~\ref{tab.enigma_vs_minion}, we see that for 12-day tracks \chtt$=65.9\%$ with the most rudimentary and optimistic detection model. The completeness drops by 6.8\%, to 59.0\%, when fill factor, fading, trailing losses, vignetting, NEO colors and light curves are applied. Accounting for the linking losses and bright source masking assumed by \citet{2017Veres_2}, we lose another 4.4\%, arriving at a final \chtt estimate of 54.6\%. 

\begin{figure}[H]
  \centering
    \plottwo{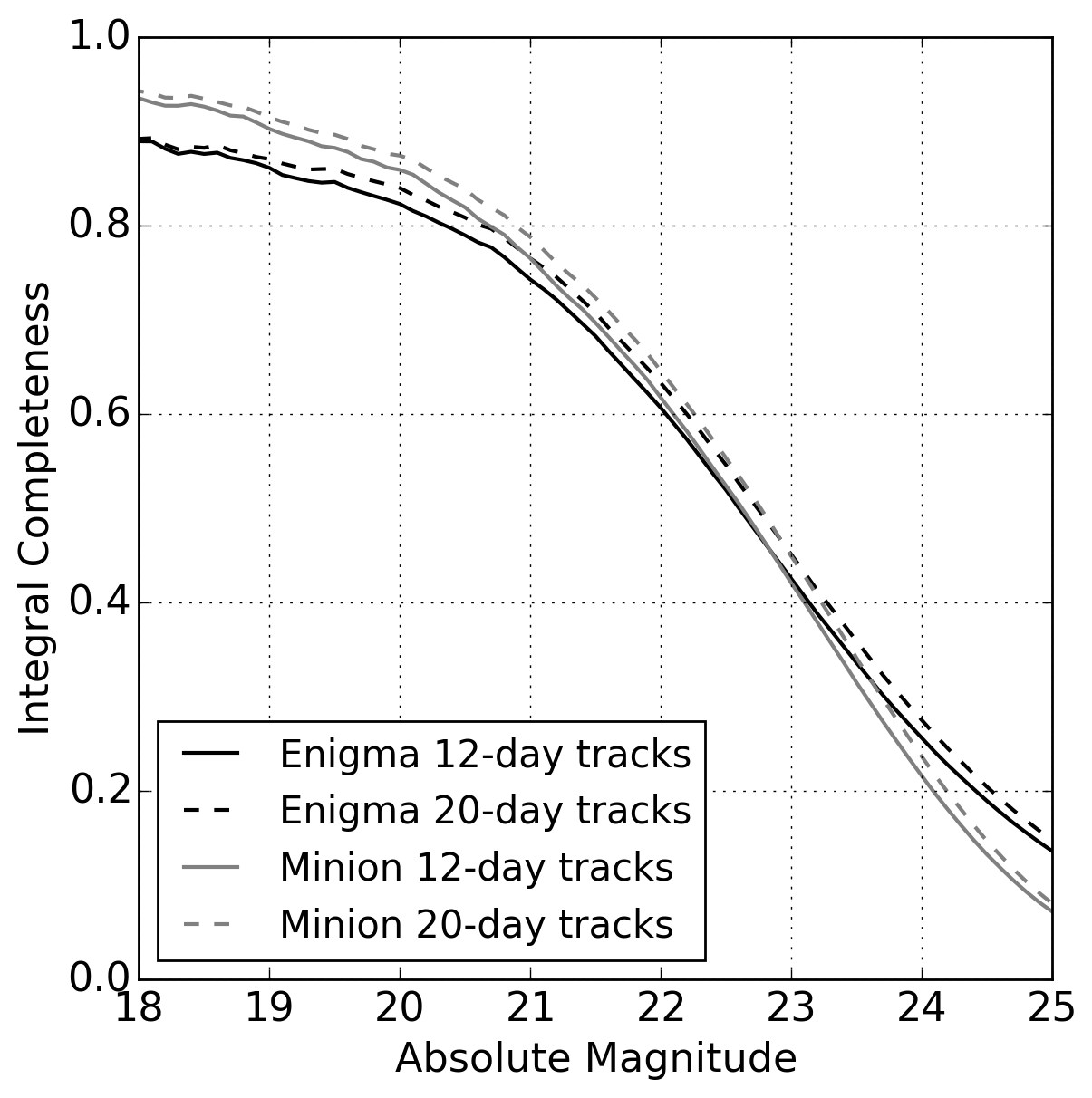}{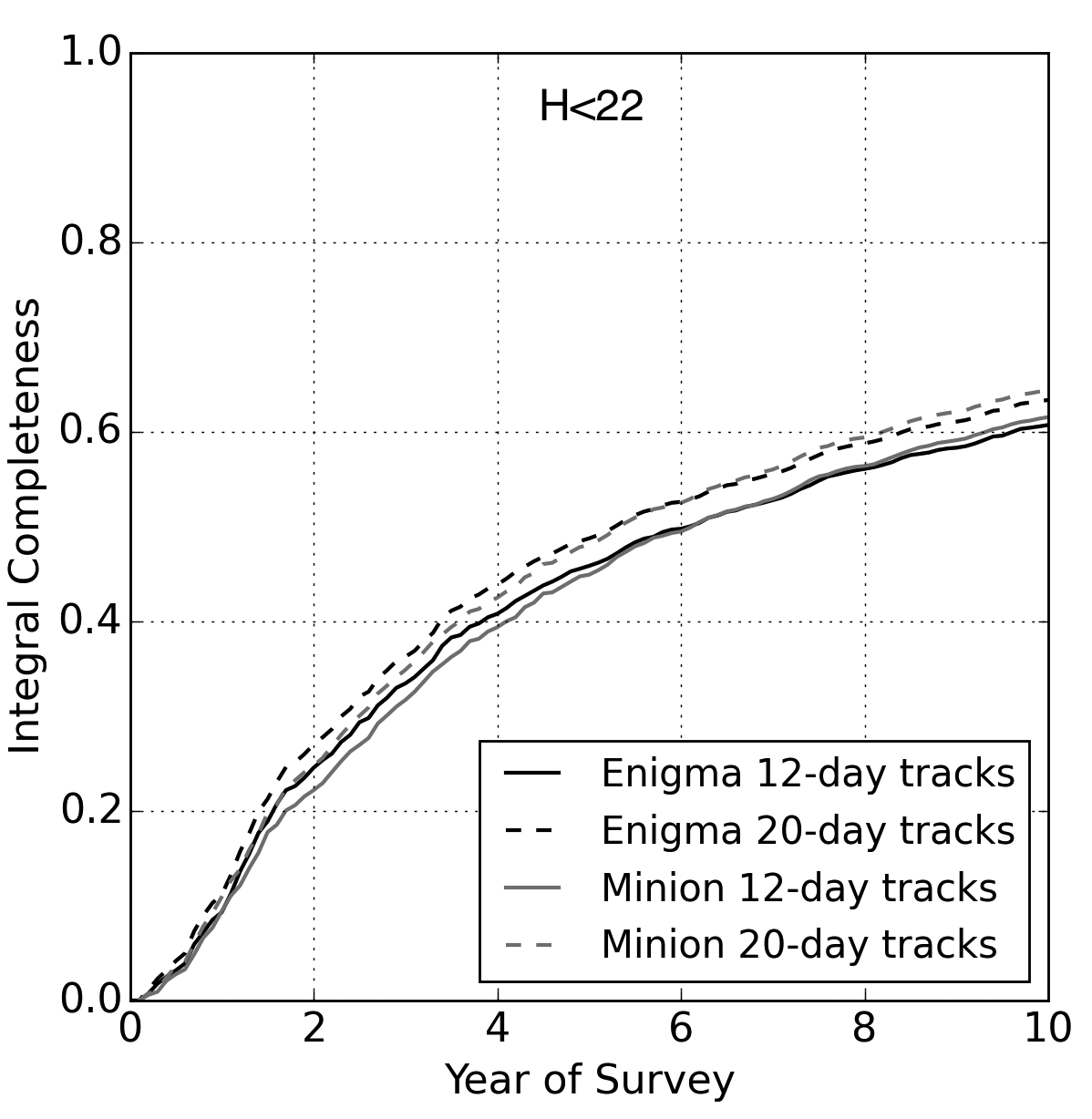}
     \caption{Integral NEO completeness for the old baseline (\enig) with Bottke's NEO model versus the new baseline (\minion) with Granvik's NEO model as a function of H (left) and time for $H<22$ (right). These results reflect full-density, NEO-only simulations with 90.8\% fill factor, fading included and detection trailing losses.}
    \label{fig.enigma_vs_minion}
\end{figure}

Comparing the \enig results with those from the \minion column, we see that there is very little difference for $C_{H<22}$ performance. with \minion only slightly better, by about 1.5\% in $C_{H<22}$ after 10 years (Figure~\ref{fig.enigma_vs_minion}). This agreement is because of the compensating factors of Granvik's steeper size distribution, which provides a $\sim 2\%$ increase in $C_{H<22}$ (Table~\ref{tab.NEOmodels}) for the \minion simulation, and the $\sim0.25\,\mathrm{mag}$ reduction in faint limit seen for \minion, which leads to a similar drop in performance. As shown in Figure~\ref{fig.enigma_vs_minion}, the effect of using two different populations is more clear at different $H$ limits. Specifically, Granvik's population is more numerous at the small end, therefore, the \minion completeness is lower for $H<25$ when compared to \enig. Also, \enig seems to be slightly more productive in the early years, while \minion catches up and passes \enig about 7 years into the survey.

\begin{sidewaystable}[htb]
\setlength\tabcolsep{1.5pt}
\caption{NEO completeness in percentage points as multiple parameters are being applied one-by-one (right column). $\Delta C$ denotes the difference between consecutive steps. The associated error in percentage points coming from uncertainty of the model is listed under the heading ``Uncert.'' The overall uncertainty on completeness is about 5\%. The \enig results used the Bottke NEO model, and the \minion results are for Granvik's NEO model.}
\begin{center}
\small
\begin{tabular}{l|cc|cc|c|l}
\tableline \tableline
 & \multicolumn{2}{|c|}{\enig}    & \multicolumn{2}{|c|}{\minion}  &  & \\
 Model Variation & $C_{H<22}$& $\Delta C$     & $C_{H<22}$& $\Delta C$   & Uncert. & Remarks\\
\hline
None                         &65.87   &          &67.66   &          &$\pm0.6$               & Assumes 100\% fill factor, hard $m_5$ cutoff, no trailing losses, etc.\\
+ fill factor (90.8\%)       &63.18   &$-2.70$   &65.07   &$-2.60$   &$^{+0.0}_{-0.8}$ & $90.8^{+0}_{-2}\,\%$ fill factor\\
+ fading                     &62.41   &$-0.76$   &64.29   &$-0.78$   &$\pm0.2$         & fading width $w=0.1\pm0.032$ mag\\
+ trailing loss              &60.30   &$-2.12$   &61.75   &$-2.54$   &$\pm$0.7         & $\pm 32\,\%$ of $\Delta m_{\mathrm{trail}}$\\
+ vignetting                 &59.91   &$-0.38$   &61.29   &$-0.46$   &$\pm$0.1         & $\pm 32\,\%$ of $\Delta m_{\mathrm{vignette}}$\\
+ colors                     &59.38   &$-0.52$   &60.67   &$-0.62$   &$\pm$0.1         & 50:50 S\&C Groups, $\pm 1\sigma$ in SDSS color indices\\
+ light curves               &59.04   &$-0.35$   &60.32   &$-0.35$   &$\pm$0.2         & $\pm 32\,\%$ of light curve amplitude\\
+ bright source removal      &58.35   &$-0.69$   &59.64   &$-0.68$   &$^{+0.2}_{-0.4}$ & $1.0^{+1.0}_{-0.5}\,\%$ masked \citep{2017Veres_2}\\
+ Linking efficiency         &54.61   &$-3.71$   &55.82   &$-3.79$   &$^{+3.3}_{-0.6}$ & $93.6^{+6}_{-1}$\,\% linking eff. \citep{2017Veres_2}\\
Population Model             &        &          &        &          &$\pm$2.0         & Granvik vs. Bottke NEO models\\
Variation in faint limit     &        &          &        &          &$^{+1.8}_{-4.4}$ & $m_5$ variation of $^{-0.10}_{+0.25}$ mag\\
\hline
OVERALL                      &54.61   &$-11.26$  &55.82   &$-11.84$  &$^{+4.3}_{-5.0}$ & $55\pm 5\%$ adequately describes final result\\
\hline
\hline
\end{tabular}
\end{center}
\label{tab.enigma_vs_minion}
\end{sidewaystable}

The completeness derived from our simulations has numerous error sources. The easiest to define are those associated with sampling. Other contributions from the various individual modeling details were derived step-by-step, and we discuss them below and summarize them in Table~\ref{tab.enigma_vs_minion}. The modeling details and our ad hoc approach for understanding their potential effect on $C_{H<22}$ are as follows:
\begin{itemize}
\item By using low-density simulations, the statistical sampling error is about 0.6\%. 
\item Using  Granvik's NEO population instead of Bottke's leads to a 2\% increase in $C_{H<22}$, and we take this as proxy for the uncertainty due to uncertainties in the NEO size distribution. 

\item Changing the fill factor by $+0/\!\!-\!2$\% leads to a shift of $+0/\!\!-\!0.8$\% in $C_{H<22}$. 
\item The assumed width for detection fading was a constant value of $w=0.1$. If this value is altered by 32\%, then the completeness error from this source is 0.2\%. 
\item Trailing losses are described by a detection loss function that is theoretical and not operationally validated. If the detection loss penalties varies by 32\% it leads to a 0.7\% error in completeness. 
\item We allow the color indices taken from SDSS and assumed in the study to vary by $1\sigma$ of their values, which affects completeness by 0.1\%.
\item We allow a 32\% variation in light curve amplitude, which leads to a 0.2\% effect on completeness.
\item For vignetting, we take a 32\% variation in the magnitude loss to obtain a 0.1\% variation in completeness.
\item Bright sources lead to a large number of false detections in difference images, which could make linking inefficient. \citet{jones2017} found that by masking areas around bright sources---about 1\% of the focal plane area---the false detection rate will be dramatically reduced. Thus we reduce the fill factor by 1\%. If this value varies by $-0.5/\!+\!1.0$\%, then the completeness error from this source is $+0.2/\!\!-\!0.4$\%. 
\item Linking efficiency is discussed at length by \citet{2017Veres_2}. Here we assume a value of $93.6^{+6}_{-1}\%$. These uncertainties cause \chtt to vary by $+3.3/\!-\!0.6$\%.
\item We take the somewhat pessimistic view that the actual operational LSST $m_5$ can vary by $+0.1/\!\!-\!0.25\,\mathrm{mag}$, which causes the completeness to vary by $+1.8/\!\!-\!4.4$\%.
\end{itemize}
We note that many of the foregoing stated uncertainties are more akin to sensitivity exercises than uncertainty estimates. In many cases we have no good statistical footing from which to infer the uncertainty in the inputs and so the corresponding uncertainty estimate relies heavily on judgement. Nonetheless, taken all together, these modeling effects lead to a $\pm5$\% uncertainty in the predicted value of $C_{H<22}$, and so, in light of this uncertainty, the difference between the old (\enig/Bottke) and new (\minion/Granvik) simulations is negligible. Therefore we report our final LSST performance result as $C_{H<22}=55\pm5\%$ after  linking losses. Here we emphasize that the stated uncertainty is not a Gaussian 1-sigma error bar, but rather reflects the possibility of modeling systematics that could compromise the result by {\em up to} 5\%.

\clearpage

\subsection{Alternatives to Nightly Pairs}

Historically, the Minor Planet Center accepts only high-reliability tracklets from observers, and its internal linking processes assume that the false tracklet rate is low. This has been a reasonable assumption for past and current major surveys, which follow a cadence that naturally returns 3--5 detections per tracklet by repeatedly returning to the same field within a span of an hour or so. This survey approach is robust against false positives because the 3--5 detections in a tracklet must all be consistent with a common linear (or nearly linear) motion, effectively eliminating the possibility that one or more false detections could contaminate a tracklet. 

LSST, on the other hand, is baselined to return only two detections per tracklet, which eliminates the possibility of checking for internal consistency among the elements of a tracklet. The result is a high rate of false tracklets that is not suitable for submission to the MPC. LSST will work past this obstacle by submitting only high-reliability three-night tracks to the MPC.

The LSST approach of obtaining nightly pairs is certainly more fragile for linking than a cadence that returns nightly triples or quads, but the fragility comes with the marked benefit of significantly increased sky coverage per night and hence a shorter return period, leading to more tracklets per observing cycle, which restores some measure of robustness and certainly leads to increased discovery rates, so long as the linking problem can be managed. If, for whatever reason, and however unlikely, LSST cannot successfully link two-detection tracklets then it could conceivably be forced to observe triples or quads to meet survey objectives. 

Here we compare the performance of the pair-wise \enig baseline survey with \enigthree and \enigfour, which are tuned to provide 3- and 4-visit cadences, respectively. For the 2-visit baseline survey, we required at least 2 detections for tracklet creation, in 3-visit baseline at least 3-detections per tracklet (``triples'') and in 4-visit cadence at least 4 detections per tracklet (``quads'').

We emphasize that the benefit of a cadence that produces triples or quads is that it eases the linkage challenge. It does not produce better orbits. Tracklets formed from three or more detections have far higher confidence than those obtained from pairs because with only two positions there is no independent corroboration of linear motion. Thus for pairs, the idea that a tracklet's detections are associated with a single moving object is a hypothesis to be tested by the linking engine, whereas with three or more detections the tracklet has a high likelihood of being real. Linking is challenging for pairs because of the high false tracklet rate and easy for triples or quads because there is no hypothesis testing involved.

In terms of orbit quality, there is no appreciable difference between orbits derived from pairs or triples or quads. Each tracklet provides a position and rate of motion on the plane of sky, with no information on plane of sky acceleration (except for very close objects, which are rare enough to be ignored in this context). The orbit quality depends primarily on how many distinct nights the object has been observed and the time interval between the first and last night.

Table~\ref{Tab.234visits} shows that visiting the same field more often per night predictably decreases the effective areal search coverage significantly, even though the alternate surveys have a similar number of fields observed in 10 years, similar limiting magnitudes, and similar inter-night survey patterns. Figure~\ref{fig.234visits} shows that all three surveys also contain some fields that are visited fewer or more than the target number of visits per night. As mentioned earlier, singleton and Deep Drilling fields are not used in our simulations.

\begin{figure}[H]
  \centering
    \includegraphics[width=0.4\textwidth]{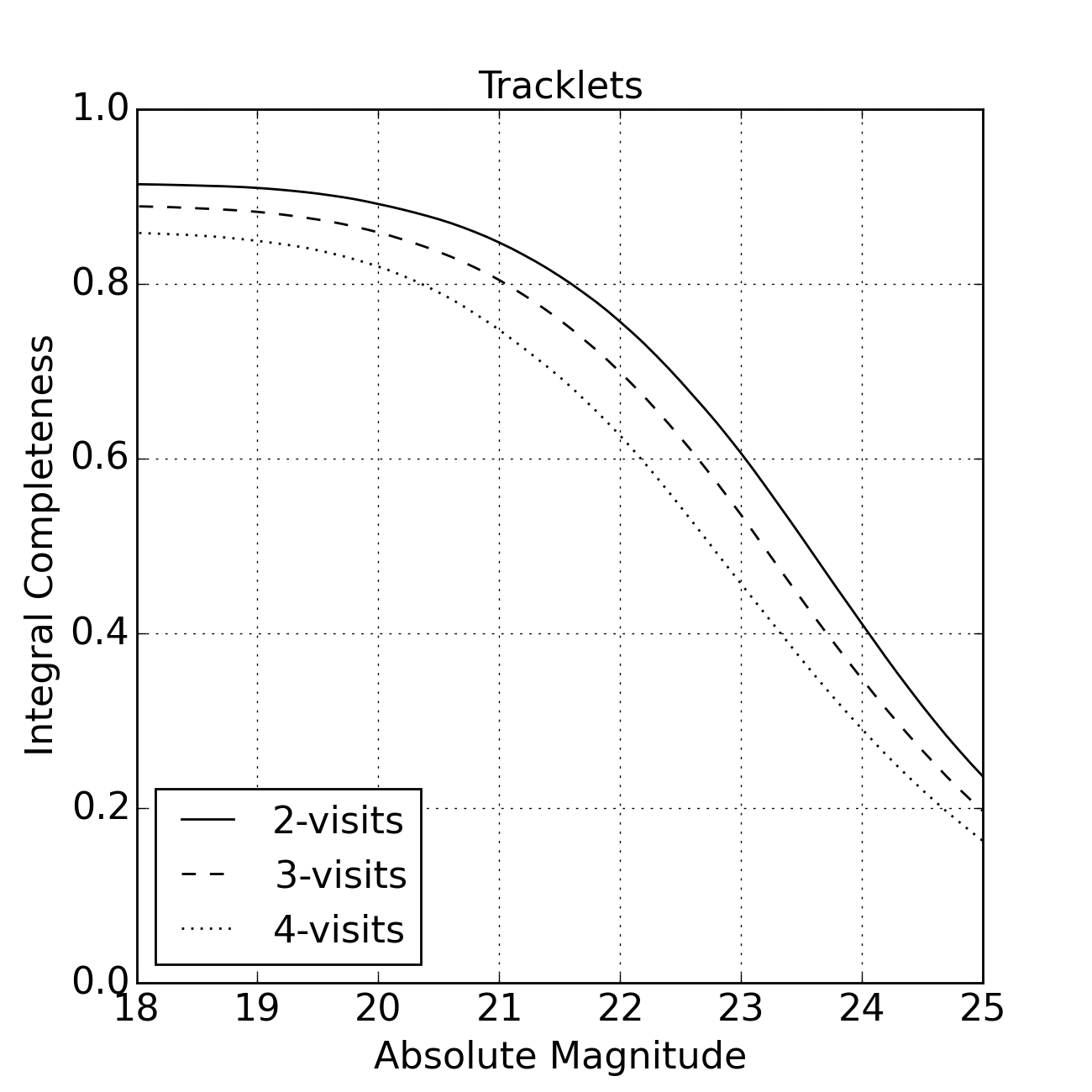}
        \includegraphics[width=0.4\textwidth]{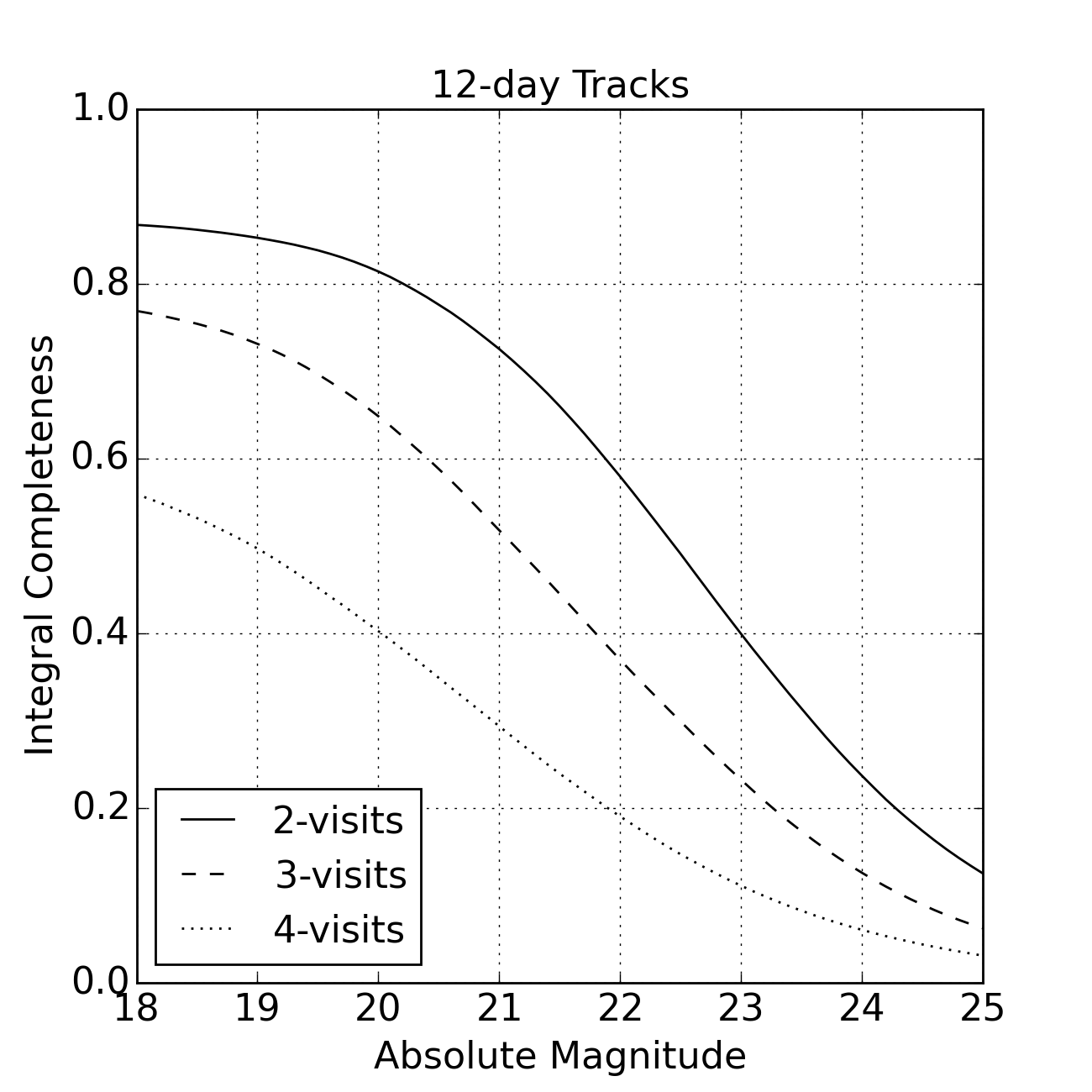}
     \caption{Integral completeness for 2, 3 and 4-visit optimized baseline surveys as a function of absolute magnitude. NEO detected at least once in a single tracklet (left) and in at least one 12-day, three-night track (right). Parameters applied: fading, fill factor=0.89, trailing loss, vignetting, colors, light curves.}
    \label{fig.234visits_comp}
\end{figure}

Figure~\ref{fig.234visits_comp} and Table~\ref{Tab.234visits_results} show the direct impact of the 3 and 4-visits cadence approach on NEO completeness. The completeness penalty is not as dramatic on the single tracklet level; however, due to reduced sky coverage the tracks are dramatically affected. Three and four-visit cadences could decrease the number of false tracklets significantly, but at the cost of a steep reduction in $C_{H<22}$. The figure shows that 3- and 4-visit cadences have a severe impact on NEO completeness in all stages, from detections to tracklets to tracks. At $H<22$, the completeness for 12-day tracks falls from $58.6$\% for nightly pairs to $36.9$\% for 3-visit cadence to $19.0$\% for 4-visit cadence. The 4-visit cadence could be improved by accepting tracklets with detections on three out of the four visits, which would provide a performance somewhere in between that tabulated (and plotted) for the 3- and 4-visit cadences. 

\begin{table}[htb]
\small
\caption{Comparison of 2, 3 and 4-visit optimized survey cadences.}
\begin{center}
\begin{tabular}{c|ccccc}
\tableline \tableline
OpSim Desig. & No. visits & No. fields & Used fields & Mean $m_5$ in V-band & Effective sky coverage ($deg^{2}$)\\
\hline
\enig&2  & 2,469,307 & 1,943,901& 23.63 & 805,128 \\
\enigthree  &3  & 2,438,990 & 1,932,391 & 23.65 & 463,636  \\
\enigfour &4  & 2,417,999 & 1,928,921 & 23.65 & 309,154 \\
\hline
\hline
\end{tabular}
\end{center}
\label{Tab.234visits}
\end{table}

\begin{table}[htb]
\small
\caption{NEO completeness in percent for 2, 3 and 4-visit optimized surveys based on the \enig baseline, with fading, fill factor equal to 0.90, trailing losses, vignetting, S+C NEO colors, light curves and bright source removal represented by 1\% penalty in fill factor.}
\begin{center}
\begin{tabular}{c|cccccc}
\tableline \tableline
& \multicolumn{6}{c}{$C_{H<22}$}\\
OpSim Desig. & No. visits &  Detection &Tracklet & 3 Tracklets & 12-day track& 20-day track\\
\hline
\enig&2  & 81.6 & 75.7 & 67.9 & 58.0 & 61.0  \\
\enigthree  &3 & 81.3 & 69.9 & 57.1& 37.0 & 43.1 \\
\enigfour &4 &  80.6 & 62.6 & 45.2 & 19.1 & 25.7\\
\hline
\hline
\end{tabular}
\end{center}
\label{Tab.234visits_results}
\end{table}

We note that triples or quads should dramatically reduce the false tracklet rate so that reliable tracks could likely be assembled from tracklets on only two nights. This approach would lead to increased completeness; however, such two night tracks would inevitably lead to weak orbits that are not likely to meet survey objectives. In particular, two-night orbits are generally uncertain enough that MBAs and NEOs cannot be distinguished in general. As discussed by \citet{2017Veres_2}, this is already an issue for some three-night orbits, but for two-night orbits it is the norm. Table~\ref{Tab.two_nighters} indicates the level of NEO completeness that would be obtained for the four-visit \enigfour survey, assuming that a minimum of either three or four detections are available for each tracklet. We did not analyze the orbit quality of two night tracks, but the challenge here is that the shortest arcs (e.g., 3 days) may be readily linked but the orbits are likely to be highly uncertain, while for the longer arcs (e.g., 12 days) the orbit may still be unacceptably weak but the linkage problem may be difficult. The numbers in Table~\ref{Tab.two_nighters} should be compared with the 12-day, 3-night tracks completeness for \enigfour in Table~\ref{Tab.234visits_results}, which is 19.1\%. Even the most optimistic scenario in Table~\ref{Tab.two_nighters} significantly underperforms compared to the baseline \enig.

\begin{table}[tb]
\small
\caption{NEO completeness $C_{H<22}$ in percent for two-night tracks based on the four-visit \enigfour survey, for a range of arc length requirements and for a requirement of either $\ge3$ and $\ge4$ detections per tracklet.}
\begin{center}
\begin{tabular}{c|cccc}
\hline
 Detections per& \multicolumn{4}{c}{Arc Length}\\[-1mm]
 tracklet       & 3 days  & 6 days & 9 days & 12 days \\
 \hline
$\ge 3$ & 23.3    & 38.3   & 46.1   & 49.9    \\
$\ge 4$ & 11.2    & 22.3   & 29.9   & 35.5    \\
\hline
\end{tabular}
\end{center}
\label{Tab.two_nighters}
\end{table}

\clearpage

\subsection{Including Prior and Ongoing Surveys}

In all of the analyses presented so far, we have implicitly assumed that no NEOs have been discovered prior to the LSST survey. However, based on current NEO discovery statistics and published population models (e.g., \citet{2002Icar..156..399B, 2015Icar..257..302H, 2016Natur.530..303G}), the population of NEOs with $H<22$ is currently already complete to a level of approximately 30\%. It is expected that this number will continue to increase until LSST becomes operational in 2022 and that at least some current or future NEO survey assets will continue to operate during the LSST mission. Therefore, some fraction of potential LSST NEO discoveries will have already been discovered by other surveys. Similarly,  some fraction of objects missed by LSST will also have already been discovered. To make an estimate of what the completeness will be after the ten-year LSST survey we must make an accounting for the contributions of other surveys.

Spacewatch \citep{2006DPS....38.5807M} was the first CCD-based NEO search program, but the era of dedicated wide-field NEA surveys began approximately 18 years ago when LINEAR \citep{2000Icar..148...21S} became operational. Since then improvements in instrumentation and techniques allowed fielding of other advanced ground based surveys like NEAT \citep{1999AJ....117.1616P}, LONEOS \citep{1999BAAS...31.1091K}, the Catalina Sky Survey and the Mt. Lemmon Survey \citep{2003DPS....35.3604L,2012DPS....4421013C} and Pan-STARRS \citep{2004SPIE.5489..667H,2010SPIE.7733E..0EK}. The space-based NEOWISE program \citep{2010AJ....140.1868W} has also made contributions to cataloging NEOs. 

We simulated past, current and presumed future NEO surveys starting 15 years in the past to 15 years in the future (2002--2032). In the simulation, LSST begins operation in 2022, about five years from this writing. Ephemerides of all objects where calculated once per day and an object was considered {\em discoverable} when all of the following criteria were met: 
\begin{itemize}
\item Apparent magnitude brighter than the detection limit $V_\mathrm{lim}$
\item Ecliptic latitude was between $\pm60\degree$
\item Geocentric opposition-centered ecliptic longitude was between $\pm90\degree$
\item Declination from $-30\degree$ to $+75\degree$ 
\item Lunar elongation $>90\degree$
\end{itemize}
 Ground-based surveys are limited by weather and cannot cover the entire sky per night; therefore, only a fraction $F_{\mathrm{disc}}$ of the objects that were discoverable according to the above criteria were added to the catalog.

Surveys have improved in time and so we slowly improve the detection model in 5-year steps:
\begin{enumerate}[I.]
\item A LINEAR-like era  from 15--10 years ago with limiting magnitude $V_\mathrm{lim}=19.5$ and $F_{\mathrm{disc}}=0.5$
\item A Catalina era with limiting magnitude of $V_\mathrm{lim}=21.0$ the next 5 years and $F_{\mathrm{disc}}=0.6$
\item For the past 5 years, Pan-STARRS1 and the Mt. Lemmon Survey have operated at the limiting magnitude of $V_\mathrm{lim}=21.5$ and $F_{\mathrm{disc}}=0.7$
\item For the next 5 years from present, we expect the limiting magnitude to be $V_\mathrm{lim}=22.0$ and $F_{\mathrm{disc}}=0.8$ to account for improvements in the combination of Pan-STARRS1, Pan-STARRS2 and both Catalina surveys. Also, the southern declination limit is extended to $-45\degree$. 
\item Starting 5 years from present we augment the previous search interval with the LSST survey and increase $F_{\mathrm{disc}}=0.9$ to account for continuing improvements for the other surveys.
\end{enumerate}

\begin{figure}[h]
  \centering
    \includegraphics[width=0.7\textwidth]{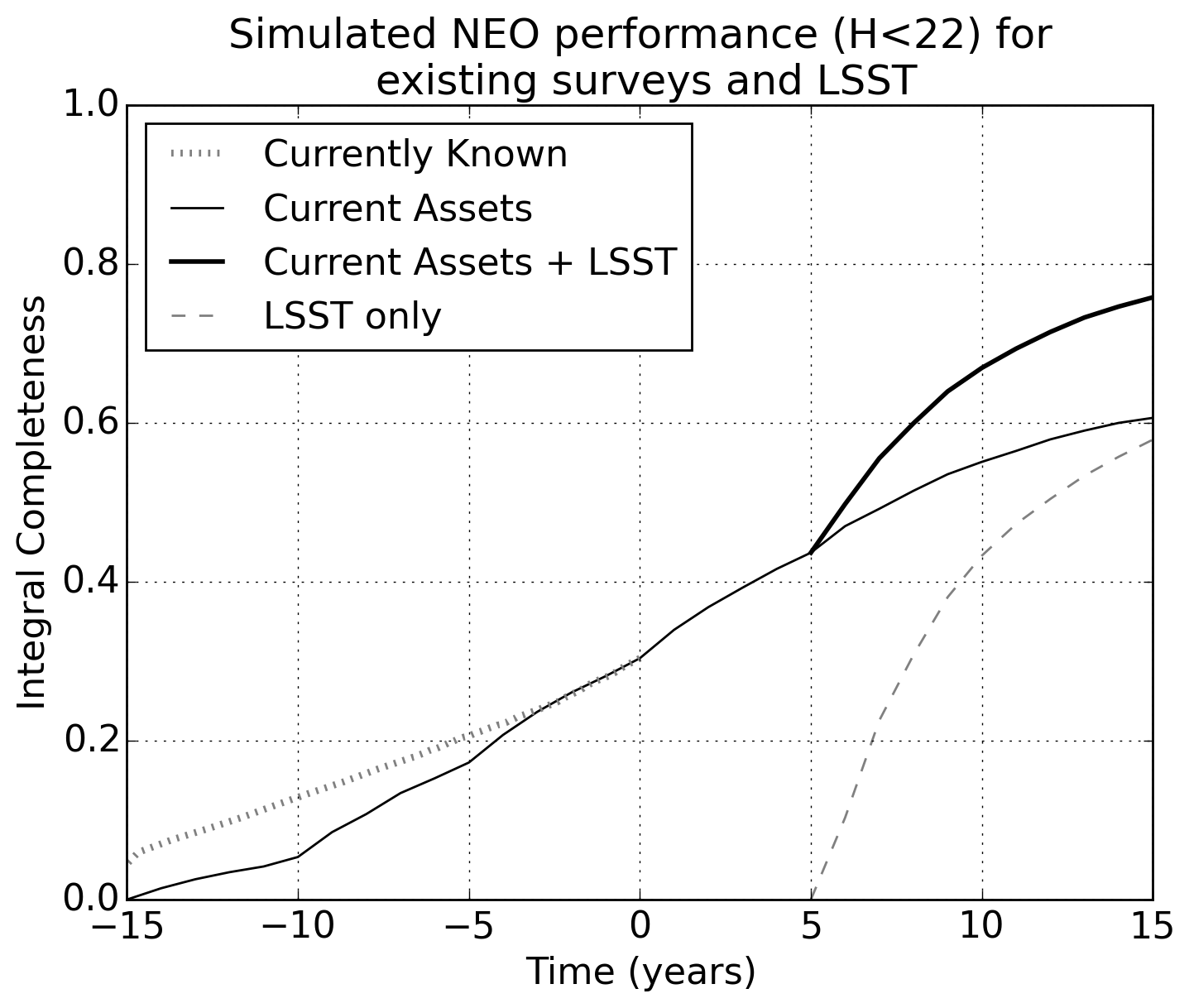}
     \caption{Estimate of integral NEO completeness ($H<22$) for past, existing and future NEO survey assets combined. The dotted curve represents the real-world NEO discovery rate, the black curve the modeled assets, the dashed curve the LSST performance as if no NEO are known at the beginning of LSST survey and the thick black curve represents the combined performance with LSST. The present date is at zero on the time axis, and LSST starts to operate in year 5. The LSST simulations are for \enig with the Bottke NEO model, 90.8\% fill factor, fading, trailing, 50/50 C \& S type asteroids and vignetting. Linking losses (bright source masking and linking efficiency) are neglected.}
    \label{fig.known}
\end{figure}

Figure~\ref{fig.known} shows the outcome of the rudimentary simulation, which is deliberately tuned to match the current estimated completeness at the current time. Our simple model predicts that 42\% of NEOs with $H<22$ will be discovered before LSST becomes operational, and that without LSST the current NEO surveys alone could achieve $C_{H<22}=61\%$ by 2032, when the LSST survey is planned to conclude. We have shown above that LSST acting alone will achieve completeness of about 58\% by itself (neglecting linking efficiency here), but when combined with past and other expected NEO search efforts, \chtt rises to 77\%. This is not a high-fidelity analysis, but it shows that the combination of LSST with other ground-based search activity will increase $C_{H<22}$ by about 20\% compared with the naive assumption of LSST starting with an empty catalog. Put another way, we project that LSST will provide a 16\% increase in $C_{H<22}$ compared with the anticipated efforts of the existing NEO search programs.

\clearpage

\subsection{Comparison with previous work}\label{ssec:compare}

Predictions of NEOs and PHAs discovery and detection rates for LSST have been done for a decade. \citet{2007IAUS..236..353I} used 1000 synthetic NEO orbits based on a large population ($D>1\km$), with assumed LSST sky coverage and revisit rate, with only a rough estimate of cadence or limiting magnitudes. \citet{2007IAUS..236..353I} predicted that LSST will detect 90\% of PHAs larger than 250\,m and  75\% of PHAs larger than 140\,m, with the optimistic estimate of 90\% completion if the cadence and survey is NEO-optimized.
\citet{Ivezic2014} used a list of LSST pointings and a higher fidelity model for the survey, using diameter-limited PHA population of 800. This analysis found 82\% PHA completeness ($D>140$m) that could be improved to 90\% with the same cadence if stretched to 12 years.

\citet{gsm16} simulated the NEO and PHA performance of LSST for the baseline \enig and 4-visit \enigfour surveys. Additionally, they discussed the discovery performance of LSST combined with existing surveys and the proposed NEOCAM space-based survey. They used the same fields as this study (except that they did not remove singletons and Deep Drilling fields), the same limiting magnitudes and a 95\% fill factor, which is higher than the currently anticipated 90.8\%.  The synthetic population was different, consisting of 20,000 NEOs down to a size threshold of 100 m. Also, \citet{gsm16} allowed creation of tracklets in a velocity range between 0.011--48 deg/day, which is significantly broader than our range (0.05--2.0 deg/day). The spectral distribution of synthetic objects had an equal balance between C and S types, however, the color transformation to V-band was derived by slightly outdated specifications from SDSS \citep{2001AJ....122.2749I}. Their simulation process was similar to ours, where detections and tracklets were assembled into lists and built into assumed tracks through post-processing. A track was $created$ when 3 tracklets were detected within 12 days, with a maximum separation of 6 days from two of them. However, for the 4-visit case, \enigfour, \citet{gsm16} required only two 4-detection tracklets within 12 days to build a track. This is a significant difference from our three tracklets on three distinct nights over at most 12 or 20-days tracks. While \citet{gsm16} were skeptical of the 2-visit cadence, given that it has never been tested operationally on a comparable survey, their 4-visit alternative only required two tracklets for a track. They also  did not include trailing losses, vignetting, fading, light curve variation, and they assumed the linking efficiency to be 100\%. \citet{gsm16} derived that for the 2-visit cadence LSST will have a 63\% completion of NEOs larger than 140\,m and for PHAs they find $C_{D>140\meter}= 62\%$. In the 4-visit cadence, they reported $C_{D>140\meter}= 59\%$ for NEOs and 58\% for PHAs. The completeness was presented with $\pm1\%$ uncertainty.  

\begin{figure}[tb]
  \centering
    \includegraphics[width=0.6\textwidth]{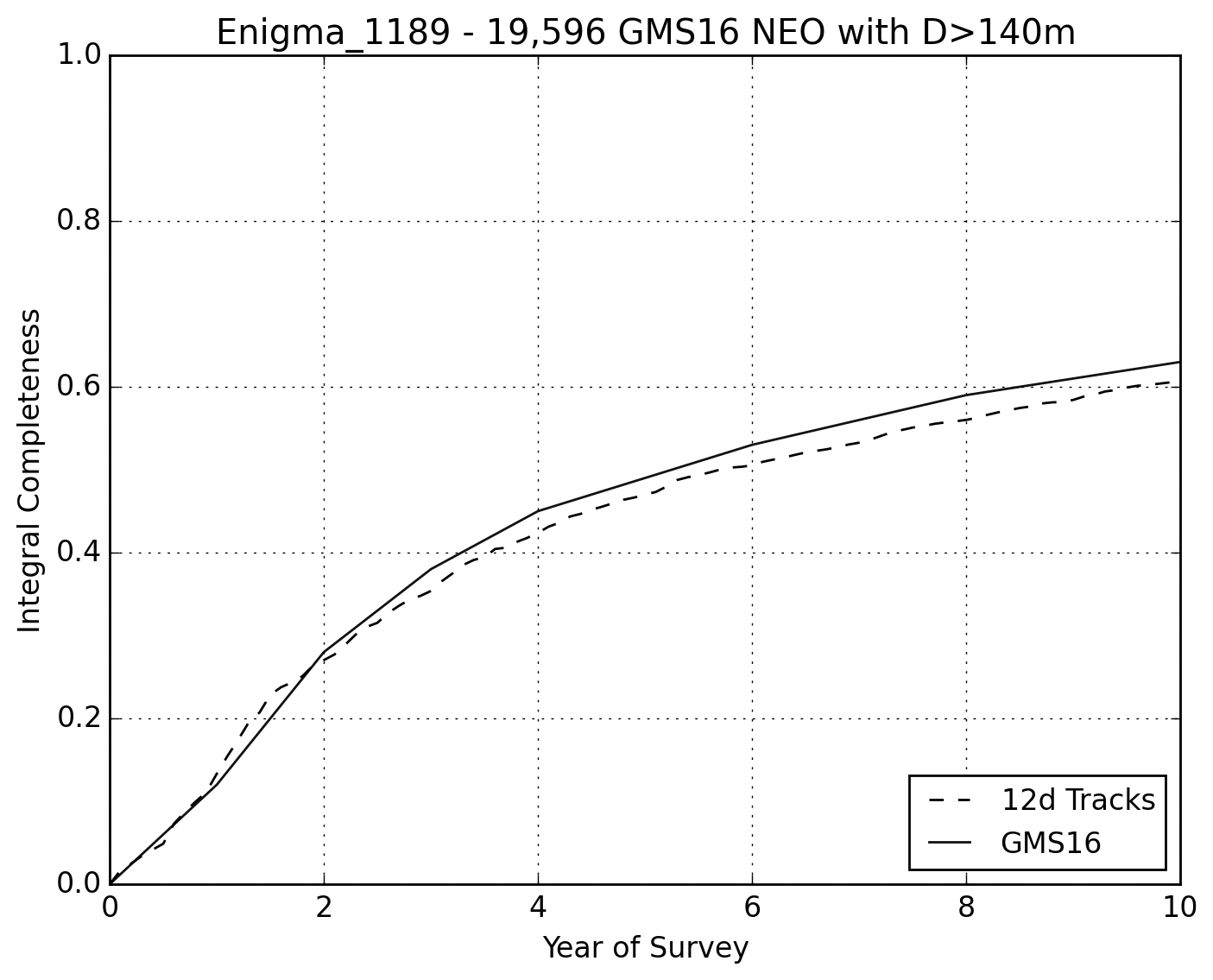}
     \caption{Comparison of 12-day tracks from the present study with the published \citet{gsm16} results. Both simulations use \enig and a compatible NEO population. The integral completeness is computed for $D>140\meter$, rather than $H<22$, which is used everywhere else in this report.}
    \label{fig.gms16_comp}
\end{figure}

To make a robust comparison with this work, we ran a simulation with a \citet{gsm16} population of NEOs (19,597) and PHAs (2,346). Our model included fading, trailing losses, vignetting and 90\% fill factor. Table~\ref{Tab.Grav} and Figure~\ref{fig.gms16_comp} present the comparison. Our results are substantially consistent with \citet{gsm16}; however, we note that our PHA completeness is slightly higher than that for NEOs, while the converse is true for \citet{gsm16}.

\begin{table}[tb]
\small
\caption{140 m integral completeness of NEAs and PHAs according to \citet{gsm16} and the present study. The columns marked ``w/others'' incorporated discoveries from other (past and future) survey activity in the completeness estimate.}
\begin{center}
\begin{tabular}{c|c|cc|cc}
\tableline \tableline
 & \multicolumn{1}{c|}{This work} & \multicolumn{4}{c}{\citet{gsm16}} \\
 & 2-det. & 2-det. & 2-det. w/others & 4-det. & 4-det. w/others \\
\hline
NEO & 62.6\% & 63\% & 67\% & 59\% & 67\% \\
PHA & 63.8\% & 62\% & 73\% & 58\% & 71\% \\
\hline
\hline
\end{tabular}
\end{center}
\label{Tab.Grav}
\end{table}

\citet{jones2017} performed a detailed computation of LSST performance using multiple parameters mentioned in this work as well as identical survey fields and patterns (\enig, \minion and \astr). Their pipeline based on MOPS was working on the detection and tracklet level, predicting tracks in post-processing of the tracklet list, similar to our low-density simulations and providing metrics within their Metrics Analysis Framework (MAF). They used an H-delimited population based on \citet{2002Icar..156..399B}. Their results are compared to ours in Table~\ref{tab.UW}, where we see an excellent match, within a percentage point, for \minion. For \astr, the UW results show a slightly increased \chtt ($\sim3\%$) compared to the present study.

There were several modeling differences between the present report and \citet{jones2017}. For example, \citet{jones2017} did not apply a rate-of-motion cut-off, which should lead to an increase in computed completeness. Also, they used all OpSim fields, without rejecting Deep Drilling and singletons, which should again lead to a slight increase in their modeled completeness. However, they considered all NEOs to be C type, and had a 1-hour restriction for a maximum tracklet duration which is half of our maximum duration. These two modeling effects could drive their estimated completeness down. Also, \citet{jones2017} used a more sophisticated model for chip and raft gaps, masking the exact locations in the focal plane, whereas our work used only a random number as a fill factor to assign a detection probability to every detection. Collectively, these model variations readily explain the slightly different NEO completeness estimates reported by \citet{jones2017}.

\begin{table*}[tb]
\small
\caption{Comparison of LSST NEO and PHA completeness \chtt in percentage for the 10-year \minion and the 15-year \astr surveys for \citet{jones2017} and this work. For consistency, tabulated results are for 15-day tracks, and so the results differ from those in Table~\ref{tab.enigma_vs_minion}. Other aspects of the results from this study are 90.8\% fill factor, with fading, trailing and the Bottke NEO population.}
\begin{center}
\begin{tabular}{c|ccc}
\tableline \tableline
NEO Pop. & Survey & \citet{jones2017} & This work \\
\hline
NEO & \minion & 60.7\%  & 61.6\%  \\
PHA & \minion & 65.6\%  & 64.9\%  \\
\hline
NEO & \astr   &  69.1\% & 65.6\%  \\ 
PHA & \astr   &  74.3\% & 71.5\%  \\
\hline
\hline
\end{tabular}
\end{center}
\label{tab.UW}
\end{table*}

\section{Conclusions}

We have demonstrated the importance of the observational constraints and input parameters on the detection of NEOs by an automated survey such as LSST. Among the numerous modeling details that we investigated, few result in a significant effect on survey performance if reasonable care is taken in developing the nominal survey model. A 2\% reduction in fill factor or a major increase in trailing losses could lead to nearly a 1\% reduction in \chtt for NEOs. We found a 2\% difference in \chtt between the Bottke and Granvik NEO models, suggesting some dependence on the population size distribution. However, the Granvik model is based on vastly superior input dataset compared to the 15-year old Bottke model. One should not expect future population models will lead to such large swings in the size distribution, and thus \chtt should be more stable in this regard. The effective, operational limiting magnitude of the LSST survey is a crucial parameter. We find that \chtt degrades by $\sim1.8\%$ for every 0.1 mag loss in sensitivity, making this the single largest source of model uncertainty in our completeness estimates.
 
 We find that \chtt$\simeq55\pm5\%$ for the baseline \minion survey operating alone and with the linking working at 93.6\% efficiency \citep{2017Veres_2}. This result assumes 12-day, 3-night tracks, and accounts for all of the foregoing modeling features and sources of error, including linking losses. The 12-day track linking is very conservative; \chtt increases by 2--3\% if the linking uses 20-day tracks in the baseline LSST cadence. 
 
 We find that \chtt for PHAs is generally 3--4\% greater than for NEOs. \citet{gsm16} reported simulations for \enig that suggest the PHA completeness is actually lower by 1\%. In contrast, \citet{jones2017} report shows a good agreement in \chtt for NEOs and PHAs in \minion simulation. For \astr, the UW results exhibit slightly larger \chtt by $\approx3$\% with respect to our results.
 
 The old and new LSST baselines (\enig and \minion, respectively) and the NEO-enhanced scenario (\astr) all provide similar NEO detection efficiencies, to within 1\%, for a 10-year survey. Surveying for longer than 10 years increases \chtt by about 2\% per year (1\% per year if other surveys are taken into consideration). The three- and four-visit optimized LSST cadences that we tested had a dramatically reduced NEO completeness relative to that obtained for the two-visit baseline cadences. This result relies on the modeling hypotheses assumed throughout this study and assumes that tracklets in three distinct nights within 12 days are required for cataloging. The performance loss associated with these alternate cadences could be significantly eased if the cataloging requirement was for tracklets on only two nights, rather than three, but it is doubtful that such two-nighters would have high enough orbit quality to meet cataloging objectives. 
 
 When LSST becomes operational in 2022, about 42\% of NEOs with $H<22$ should have been discovered with the current assets. Without LSST, current assets could discover 61\% of the catalog during the LSST era. With LSST and other surveys combined, \chtt should reach 77\% by the end of 10-year mission in 2032, assuming 12-day tracks with \enig. 
 
 Assembling the foregoing completeness results, including the contribution of other surveys, the post-LSST \chtt should reach 80\% for PHAs in ten years, and slightly more if 20-day tracks are linked. In our judgement, the $H < 22$ PHA catalog is likely to approach $\sim85\%$ completeness, but probably not 90\%, after 12 years of LSST operation.

\acknowledgements
\begin{center}
{\em Acknowledgments} 
\end{center}

The Moving Object Processing System was crucial to the successful completion of this work. This tool is the product of a massive development effort involving many contributors, whom we do not list here but may be found in the author list of \citet{2013PASP..125..357D}. We thank Larry Denneau (IfA, University of Hawaii) for his tremendous support in installing and running the MOPS software.

This study benefited from extensive interactions with Zeljko Ivezic, Lynne Jones and Mario Juric, all from the University of Washington. As members of the LSST project, they provided vital guidance in understanding the performance and operation of LSST. They also provided important insight into the expected interpretation and reliability of LSST data. They ensured that some of the OpSim runs required to fulfill study objectives were generated by LSST. 

Mikael Granvik (Univ. Helsinki) kindly provided an early version of the \citet{2016Natur.530..303G} NEO population model, which was used extensively in this work.

This  research  was  conducted  at  the  Jet  Propulsion  Laboratory, California Institute of Technology, under a contract with the National Aeronautics and Space Administration.

\vspace{.3cm}

\noindent\copyright\ Copyright 2017 California Institute of Technology. Government sponsorship acknowledged.


\begin{thebibliography}

\bibitem[Araujo-Hauck et al.(2016)]{2016SPIE.9906E..0LA} Araujo-Hauck, C., Sebag, J., Liang, M., et al.\ 2016, \procspie, 9906, 99060L 

\bibitem[Bottke et al.(2002)]{2002Icar..156..399B} Bottke, W.~F., Morbidelli, A., Jedicke, R., et al.\ 2002, \icarus, 156, 399 

\bibitem[Carvano et al.(2010)]{2010A&A...510A..43C} Carvano, J.~M., Hasselmann, P.~H., Lazzaro, D., \& Moth{\'e}-Diniz, T.\ 2010, \aap, 510, A43 

\bibitem[Chance \& Kurucz(2010)]{chance_2010} Chance, K., \& Kurucz, R.~L.\ 2010, \jqsrt, 111, 1289

\bibitem[Christensen et al.(2012)]{2012DPS....4421013C} Christensen, E., Larson, S., Boattini, A., et al.\ 2012, AAS/Division for Planetary Sciences Meeting Abstracts, 44, 210.13 

\bibitem[Delgado et al.(2014)]{2014SPIE.9150E..15D} Delgado, F., Saha, A., Chandrasekharan, S., et al.\ 2014, \procspie, 9150, 915015 

\bibitem[Denneau et al.(2013)]{2013PASP..125..357D} Denneau, L., Jedicke, R., Grav, T., et al.\ 2013, \pasp, 125, 357 

\bibitem[DeMeo et al.(2009)]{demeo_pds_2009} DeMeo, F., Binzel, R.~P., Slivan, S.~M., \& Bus, S.~J.\ 2009, NASA Planetary Data System, 114,

\bibitem[DeMeo \& Carry(2014)]{2014Natur.505..629D} DeMeo, F.~E., \& Carry, B.\ 2014, \nat, 505, 629 

\bibitem[Flaugher et al.(2015)]{2015AJ....150..150F} Flaugher, B., Diehl, H.~T., Honscheid, K., et al.\ 2015, \aj, 150, 150 

\bibitem[Granvik et al.(2016)]{2016Natur.530..303G} Granvik, M., Morbidelli, A., Jedicke, R., et al.\ 2016, \nat, 530, 303 

\bibitem[Grav et al.(2011)]{2011PASP..123..423G} Grav, T., Jedicke, R., Denneau, L., et al.\ 2011, \pasp, 123, 423 

\bibitem[Grav et al.(2016)]{gsm16} Grav, T., Mainzer, A.~K., \& Spahr, T.\ 2016, \aj, 151, 172 

\bibitem[Harris \& D'Abramo(2015)]{2015Icar..257..302H} Harris, A.~W., \& D'Abramo, G.\ 2015, \icarus, 257, 302 

\bibitem[Hodapp et al.(2004)]{2004SPIE.5489..667H} Hodapp, K.~W., Siegmund, W.~A., Kaiser, N., et al.\ 2004, \procspie, 5489, 667 

\bibitem[Ivezi{\'c} et al.(2001)]{2001AJ....122.2749I} Ivezi{\'c}, {\v Z}., Tabachnik, S., Rafikov, R., et al.\ 2001, \aj, 122, 2749 

\bibitem[Ivezi{\'c} et al.(2007)]{2007IAUS..236..353I} Ivezi{\'c}, {\v Z}., Tyson, J.~A., Juri{\'c}, M., et al.\ 2007, Near Earth Objects, our Celestial Neighbors: Opportunity and Risk, 236, 353 

\bibitem[Ivezic et al.(2009)]{2009AAS...21346003I} Ivezic, Z., Tyson, J.~A., Axelrod, T., et al.\ 2009, Bulletin of the American Astronomical Society, 41, 460.03 

\bibitem[Ivezi{\'c} et al.(2014)]{Ivezic2014} Ivezi{\'c}, {\v Z}., Tyson, J.~A., Abel, B. et al.\ 2014, arXiv:0805.2366

\bibitem[Jones et al.(2016)]{2016IAUS..318..282J} Jones, R.~L., Juri{\'c}, M., \& Ivezi{\'c}, {\v Z}.\ 2016, IAU Symposium, 318, 282 

\bibitem[Jones et al.(2017)]{jones2017} Jones, R.~L., Slater, C. Moeyens, J., Allen, L., Juri{\'c}, M., \& Ivezi{\'c}, {\v Z}.\ 2017, ``The Large Synoptic Survey Telescope as a Near-Earth Object Discovery Machine,'' submitted to Icarus.

\bibitem[Kaiser et al.(2002)]{2002SPIE.4836..154K} Kaiser, N., Aussel, H., Burke, B.~E., et al.\ 2002, \procspie, 4836, 154 

\bibitem[Kaiser et al.(2010)]{2010SPIE.7733E..0EK} Kaiser, N., Burgett, W., Chambers, K., et al.\ 2010, \procspie, 7733, 77330E 

\bibitem[Koehn \& Bowell(1999)]{1999BAAS...31.1091K} Koehn, B.~W., \& Bowell, E.\ 1999, \baas, 31, 12.02 

\bibitem[Larson et al.(2003)]{2003DPS....35.3604L} Larson, S., Beshore, E., Hill, R., et al.\ 2003, Bulletin of the American Astronomical Society, 35, 36.04 


\bibitem[Mainzer et al.(2011)]{2011ApJ...743..156M} Mainzer, A., Grav, T., Bauer, J., et al.\ 2011, \apj, 743, 156 

\bibitem[Mainzer et al.(2015)]{2015DPS....4730801M} Mainzer, A.~K., Wright, E.~L., Bauer, J., et al.\ 2015, AAS/Division for Planetary Sciences Meeting Abstracts, 47, 308.01 

\bibitem[Masiero et al.(2009)]{2009Icar..204..145M} Masiero, J., Jedicke, R., {\v D}urech, J., et al.\ 2009, \icarus, 204, 145 

\bibitem[McMillan \& Spacewatch Team(2006)]{2006DPS....38.5807M} McMillan, R.~S., \& Spacewatch Team 2006, Bulletin of the American Astronomical Society, 38, 58.07 

\bibitem[Pravdo et al.(1999)]{1999AJ....117.1616P} Pravdo, S.~H., Rabinowitz, D.~L., Helin, E.~F., et al.\ 1999, \aj, 117, 1616 

\bibitem[Schunov{\'a}-Lilly et al.(2017)]{2017Icar..284..114S} Schunov{\'a}-Lilly, E., Jedicke, R., Vere{\v s}, P., Denneau, L., \& Wainscoat, R.~J.\ 2017, \icarus, 284, 114 

\bibitem[Smith et al.(2014)]{2014SPIE.9147E..79S} Smith, R.~M., Dekany, R.~G., Bebek, C., et al.\ 2014, \procspie, 9147, 914779 

\bibitem[Stokes et al.(2000)]{2000Icar..148...21S} Stokes, G.~H., Evans, J.~B., Viggh, H.~E.~M., Shelly, F.~C., \& Pearce, E.~C.\ 2000, \icarus, 148, 21 

\bibitem[Stuart \& Binzel(2004)]{2004Icar..170..295S} Stuart, J.~S., \& Binzel, R.~P.\ 2004, \icarus, 170, 295 

\bibitem[Vere{\v s} et al.(2012)]{2012PASP..124.1197V} Vere{\v s}, P., Jedicke, R., Denneau, L., et al.\ 2012, \pasp, 124, 1197 

\bibitem[Vere{\v s} et al.(2015)]{2015Icar..261...34V} Vere{\v s}, P., Jedicke, R., Fitzsimmons, A., et al.\ 2015, \icarus, 261, 34 

\bibitem[Vere{\v s} \& Chesley(2017)]{2017Veres_2} Vere{\v s}, P. \& Chesley, S.R., \ 2017, \aj, submitted 

\bibitem[Warner et al.(2009)]{2009Icar..202..134W} Warner, B.~D., Harris, A.~W., \& Pravec, P.\ 2009, \icarus, 202, 134 

\bibitem[Wright et al.(2010)]{2010AJ....140.1868W} Wright, E.~L., Eisenhardt, P.~R.~M., Mainzer, A.~K., et al.\ 2010, \aj, 140, 1868-1881 

\bibitem[Xin et al.(2015)]{2015ApOpt..54.9045X} Xin, B., Claver, C., Liang, M., et al.\ 2015, \ao, 54, 9045 

\end{thebibliography}
\end{document}